\documentstyle[aps,prb,epsfig,multicol]{revtex}

\title{Conductance fluctuations in diffusive rings: \\
       Berry phase effects and criteria for adiabaticity }

\author{Hans-Andreas Engel \thanks{email: Hans-A.Engel@unibas.ch}
    and Daniel Loss  \thanks{email: Daniel.Loss@unibas.ch} }

\address{Department of Physics and Astronomy, University of Basel,\\
 Klingelbergstrasse 82, CH-4056 Basel, Switzerland}

\date{\today}

\renewcommand{\vec}[1]{{\mathbf #1}}
\newcommand{\lsgtag}{{\rm LSG}}
\newcommand{\adtag}{{\rm ad}}
\newcommand{\homtag}{{\rm hom}}
\newcommand{\omegabohr}{{\omega_{\rm B}}}
\newcommand{\omegabohrTilde}{{\omega_{\tilde{\rm B}}}}

\newcommand{\mubohr}{{\mu_{\rm B}}}
\newcommand{\vfermi}{{v_{\rm F}}}

\newcommand{\expectation}[1]{\left\langle #1 \right\rangle}

\newcommand{\ket}[1]{\left| #1 \right\rangle}
\newcommand{\braket}[2]{\left\langle #1 | #2 \right\rangle}
\newcommand{\braopket}[3]{\left\langle #1 \left| #2 \right| #3 \right\rangle}
\newcommand{\smbraopket}[3]{\langle #1 | #2 | #3 \rangle}
\newcommand{\cdket}[1]{\ket{#1}_{C/D}}
\newcommand{\bsigma}{\mbox{\boldmath{$\sigma$}}}
\newcommand{\deltaCD}[1]{ {\delta^{C/D}_{#1 \alpha}(j)} }
\newcommand{\deltaCDSq}[1]{ \deltaCD{#1}^2 }
\newcommand{\tDeph}{w}
\newcommand{\gammatilde}{\tilde{\gamma}^{C/D}_{\eta,\,\tilde{\eta}}}
\newcommand{\gammatildez}{\tilde{\gamma}^{C/D}_{0,\,\tilde{\eta}}}
\newcommand{\gammalsg}{\gamma^{C/D}_\lsgtag}
\newcommand{\dgUCF}{\delta g^{(2)}}
\newcommand{\dgMC}{\delta g}

\newcommand{\miAl}{a}
\newcommand{\miDel}{c}
\newcommand{\bnm}{b_{nm}}

\newcommand{\SO}{{\scriptscriptstyle\mathit SO}}

\newcommand{\half}{{\textstyle \frac{1}{2}}}

\newcommand{\vecIndex}[1]{{\raisebox{-0.5ex}{$\scriptstyle \vec{#1}$}}}

\newcommand{\cosx}[1]{\cos{\textstyle \frac{2 \pi #1 x}{L}}}
\newcommand{\sinx}[1]{\sin{\textstyle \frac{2 \pi #1 x}{L}}}
\newcommand{\emcsep}{ \noindent\setlength{\unitlength}{1mm}
 \begin{picture}(86,2)\put(0,1.5){\line(1,0){84.5}}\put(84.5,1.5){\line(0,1){2}}
 \end{picture} }
\newcommand{\bmcsep}{ \noindent\setlength{\unitlength}{1mm}
 \begin{picture}(170,4)\put(92.5,2){\line(1,0){85}}\put(92.5,2){\line(0,-1){2}}
 \end{picture} }

\newcommand{\figwidth}{\ifpreprintsty400pt\else8.6cm\fi}

\begin{document}

\maketitle

\begin{abstract}
We study Berry phase effects on conductance properties of diffusive 
mesoscopic conductors,  which are caused by an electron spin moving through an
orientationally inhomogeneous  magnetic field.
Extending previous work, we start with an exact, 
 i.e.\ not assuming adiabaticity,
 calculation of the universal conductance fluctuations
 in a diffusive ring within the weak localization regime,
based on a differential equation which we derive
 for the diffuson in the presence of Zeeman coupling to a magnetic field
 texture.
We calculate the field strength required for adiabaticity
 and show that this strength is reduced by the diffusive motion.
We demonstrate that not only the phases but also the
 amplitudes of the $h/2e$ Aharonov-Bohm oscillations are
strongly affected by the Berry phase. In particular, we show
that these amplitudes are completely suppressed
at certain magic tilt angles of the
external fields, and thereby provide a useful criterion 
for experimental searches.
We also discuss Berry phase-like effects resulting from spin-orbit
 interaction in diffusive conductors
 and derive exact formulas for both magnetoconductance
 and conductance fluctuations.
We discuss the power spectra of
 the magnetoconductance and the conductance fluctuations
 for inhomogeneous magnetic fields and for spin-orbit interaction. 
\end{abstract}


\ifpreprintsty\else\begin{multicols}{2}\fi

\section{Introduction and Overview}

Since its discovery, the Berry phase\cite{berry} has been a
 subject of continued interest.
As this geometrical phase emerges from the very basic laws of
 quantum mechanics, it has implications for a broad range of
 physical systems.\cite{review}
Even though the Berry phase has been observed in single-particle experiments,
 its manifestation in condensed matter systems is still under investigation.
Some settings were proposed, \cite{LGB1,LGB2,LGlong,LSG,lygellerDWalls}
 where the Berry phase, resulting from the
 motion of a spin-carrying particle through an inhomogeneous magnetic 
 field $\vec{B}(\vec{x})$,
 can be observed in mesoscopic structures.
The expected effects are measurable as persistent
 currents \cite{LGB1,LGlong,LGexp}
 as well as in the 
 magnetoconductance\cite{LGB2,LSG,Stern,langen,LSG99}
 and the universal conductance fluctuations 
 (UCFs).\cite{LGB2,LSG}
The first experiments reporting such effects were realized with
 semiconductor structures:
  the conductance was investigated in an InAs sample,\cite{morpurgo}
  where the Berry phase can emerge through the 
  Rashba effect\cite{aronovLygeller},
  in a very similar way as produced by an inhomogeneous field.
Magnetoconductance measurements were performed 
  where a ferromagnetic dot, placed slightly above
  a GaAs sample, produced an inhomogeneous field.\cite{ye}
Measurements on metallic systems also showed effects,
 which have been explained in terms of the 
 Berry phase.\cite{jacobs,mohantyUnp,haeussler}
Further experiments on metallic systems are in progress.\cite{nunes}
An additional scenario was proposed, 
 where domain walls of mesoscopic ferromagnets 
 lead to a Berry phase.\cite{lygellerDWalls}

During orbital motion in a magnetic field, 
 a spin acquires a Berry phase in a similar way
 as a charge collects an Aharonov-Bohm phase.
Thus, these two phases lead to similar implications for interference
 phenomena in mesoscopic samples.
However, in the first case the phase originates from the change in local field
 direction, whereas in the second case it results from an enclosed magnetic flux.
As these field properties can be varied individually, the interplay of
 the two phases yields a rich variety of behaviour.
These quantum phases are distinguished by another important difference:
 while Aharonov-Bohm effects appear for arbitrarily small magnitudes $B$ of
 the magnetic field, 
 Berry phase effects appear to their full extent only in the adiabatic
 limit, i.e.\ for large enough fields (specified below).
The physical situation required for this limit to be satisfied
 can be pictured \cite{LSG,LSG99} as a spin which must complete many 
 precessions $\omegabohr t_o /2\pi$ around
 the local magnetic field,
 while it moves during a time $t_o$ 
 through a region of size $\ell_B$ over which the direction of
 the field changes significantly.
Here we have introduced the Bohr frequency  $\omegabohr=g \mubohr B/2\hbar$,
 where $g$ is the Land\'e g-factor and $\mubohr$ is the Bohr magneton.
For ballistic motion as it occurs in clean semiconductors,
 one has $\vfermi t_o \sim \ell_B$ and there is general consensus
 about the criterion for adiabaticity, 
 i.e.\ $\omegabohr \ell_B/\vfermi \gg 2\pi$,
 with $\vfermi$ being the Fermi velocity.
However, for diffusive systems there were recently some 
 discussions\cite{Stern,LSG,langen,LSG99} 
 whether $t_o$ can be correctly set 
 as the diffusion time $t_d = \ell_B^2/D$
 or if one should replace it by the elastic scattering time $\tau$.
The first criterion is more optimistic, in the sense that much
 lower field magnitudes are required to reach adiabaticity, as
 due to diffusive motion the electrons effectively move more slowly
 (compared to the ballistic motion)
 through the changing magnetic field and thus have more time
 to adjust their spins to the local field orientation.
For magnetoconductance quantitative values for the required 
 field magnitudes have been obtained.\cite{LSG99}
 Solving the special case of a cylindrically symmetrical texture exactly, it
 was confirmed \cite{LSG99} that 
the more favorable criterion is indeed sufficient.
We remark that, if the ballistic criterion was appropriate
 for diffusive systems, 
 the large fields required for adiabaticity
 would imply a strong curvature of the semiclassical trajectories
 (apart from the case of very large $g$ factors).
This curvature in turn is in conflict with 
 the approximation of the orbital motion
 by its zero-field value and therefore
 an approach beyond weak localization theory would be required 
 for a self-consistent theory.
At this point it should also be noted 
 that Berry phase effects occur even if the adiabatic limit is not fully reached;
 there is no sharp cutoff where the Berry phase
 disappears completely. 
Thus, calculations without assuming adiabaticity
 are very desirable, as they can be used 
 to study how the Berry phase effects gradually emerge
 while the magnetic field is increased from low to adiabatic strengths.
The adiabatic limit can still be taken at the end of the calculation,
 so the formal appearance of the Berry phase and 
 the associated dephasing\cite{LSG99} can be identified.

%SO
Besides having a spin following the direction of an inhomogeneous
 external field, 
 there is another scenario which produces a Berry phase: 
 spin-orbit coupling.\cite{aronovLygeller}
If an electron moves through an electrical field perpendicular to
 the ring plane, an effective
 magnetic field, which is produced in the rest frame of the electron,
 couples to the electron spin.
As this effective field is in radial direction of the ring and
 perpendicular to the direction of motion, 
 the field rotates while the electron moves around the ring
 and can therefore produce a Berry phase.
By switching on, in addition, an external magnetic field, an arbitrary
 tilt angle of the total effective field can be realized
 and so this Berry phase can be tuned.
For ballistic motion, the Berry phase manifests itself in precisely the
 same way \cite{aronovLygeller} as in the case
 with an inhomogeneous external magnetic field.\cite{LGB1,LGB2,LGlong} 
However, for diffusive motion the situation becomes more complicated,
 as the change of the direction of motion of the 
 electron due to a elastic scattering event
 abruptly changes the effective field direction.
Now the picture of a spin, moving adiabatically through a slowly 
 varying field, is no longer valid and needs to be modified.
This leads to a new physical situation which has to be 
 considered separately from the situation with inhomogeneous fields.

%Outline
The outline of this paper is as follows.
In Sec.~\ref{secUCF} we study the 
 conductance fluctuations $\dgUCF$
 of quasi-1D diffusive rings in
 inhomogeneous magnetic fields.
While $\dgUCF$ has
 already been calculated within the adiabatic approximation,\cite{LSG}
 i.e.\ for strong magnetic fields,
 the behavior outside the adiabatic limit and the 
 influence of inhomogeneous fields on dephasing 
 were not dicussed so far.
We address these issues in the present work,
 starting in Sec.~\ref{ssecExactUCF} with a calculation of 
 an exact expression for $\dgUCF$
 (i.e.\ allowing arbitrarily small field magnitudes)
 for a special texture [see Eq.~(\ref{eqnField})] of the magnetic field.
In this process we derive a new form of the diffuson 
 differential equation, which includes inhomogeneous magnetic fields.
We evaluate the adiabatic limit of the UCFs, $\dgUCF_\adtag$,
 in Sec.~\ref{ssecUCFAd}
 and compare our results with those derived in previous work.\cite{LSG}
Further, we investigate in Sec.~\ref{ssecFiniteTemperatures}
 the finite temperature behavior of the conductance fluctuations.
In Sec.~\ref{secBerryAd} the effects of the Berry phase on the UCFs and 
 their dependence on magnetic field strengths are discussed in detail.
We identify in Sec.~\ref{ssecQual} 
 a new effect of the Berry phase by showing that the 
 amplitudes of the $h/2e$ Aharonov-Bohm oscillations depend
 directly on the value of the Berry phase.
In particular, 
 we find some magic tilt angles of the magnetic field, 
 where these Aharonov-Bohm oscillations are completely suppressed.
This effect provides a tool for experimental searches
 of the Berry phase.
We use this observation to illustrate the gradually appearing effects
of the Berry phase
 for increasing field strengths 
 and thus give a direct demonstration of the onset of adiabaticity.
Then, in Sec.~\ref{ssecQuant},
 we give quantitative values of the fields strengths
 needed for reaching adiabaticity.
We show that 
 the criterion for adiabaticity
 is less stringent for diffusive than for ballistic motion.
An exact evaluation of magnetoconductance $\dgMC_\SO$
 and conductance fluctuations $\dgUCF_\SO$
 in the presence of spin-orbit coupling and homogeneous magnetic fields
 is given in Sec.~\ref{secSOExact}.
These results show how  the amplitudes of the Aharonov-Bohm oscillations
 in $\dgUCF_\SO$ 
 depend non-monotoneously on the direction of an effective field,
 similarly as it is the case for inhomogeneous magnetic fields.
In Sec.~\ref{secSplitLimits}
 we show how frequency shift of the Aharonov-Bohm oscillations
 appear in $\dgMC$ and $\dgUCF$ caused by the Berry phase.
We then point out in Sec.~\ref{ssecFreqShiftHomFields} that
 the Zeeman term can also produce frequency shifts even in the case
 of homogeneous fields.
In Sec.~\ref{ssecPSTechniques}
 we plot and discuss the exact expressions for $\dgMC$ and $\dgUCF$
 for inhomogeneous fields and for spin-orbit coupling
 as well as the corresponding power spectra.
In three appendices we provide details of our calculations.

\section{Conductance Fluctuations}
\label{secUCF}
As foundation for further discussions of Berry phase effects and adiabaticity, 
 we will first calculate the conductance fluctuations $\dgUCF$
 in the weak-localization regime.
 To motivate the analysis of the conductance fluctuations,
 we would like to emphasize the advantage of studying the UCFs instead
 of the magnetoconductance.
The latter quantity has only contributions from the cooperon, which
 are suppressed by moderately large magnetic fields penetrating the ring 
 arms.\cite{aronov}
This suppression is in direct competition with 
 the requirement of having large fields
 to satisfy adiabaticity.
In contrast, the conductance fluctuations also have contributions
 from the diffuson, which
 is only sensitive to the {\it difference} of the two magnetic fields,
 for which the conductance correlator is considered. 
Therefore, if both fields are taken of similar magnitude,
 Aharonov-Bohm oscillations and Berry phase effects
 in the UCFs will still be visible at
 high magnetic fields
 where the adiabatic criterion is certainly satisfied.

\subsection{Exact solution}
\label{ssecExactUCF}

We shall concentrate on rings with circumference $L$ and study 
the conductance-conductance correlator
$\dgUCF ( \vec{B},\vec{\tilde{B}} ) = 
   \expectation{g_\vecIndex{B}  g_\vecIndex{\tilde{B}} } -
   \expectation{g_\vecIndex{B}} \expectation{g_\vecIndex{\tilde{B}} }$, 
 where we have two different magnetic fields 
   $\vec{B}$ and $\vec{\tilde{B}}$.
We consider a special texture\cite{LGlong,langen,LSG99} 
 for which we obtain exact results
 (i.e.\ without making the adiabatic assumption of strong magnetic fields).
We assume the magnetic fields to be applied 
 in such a way that they wind $f$ times around the $z$-axis
 in one turn around the ring,
 with tilt angles $\eta$, $\tilde{\eta}$,
 see Fig.~\ref{figRing}.
The position along the direction of the ring 
 is described by the coordinate $x$, 
 varying from $0$ to $L$,
 so the special texture of the magnetic field is expressed as
\begin{eqnarray}
\label{eqnField}
 \vec{B} &=& B \, \vec{n}  
 \\ \nonumber & %noPP
 =
 & %noPP
 B\,(\sin{\eta}\, \cos{\textstyle (\frac{2 \pi f x}{L}\!+\!\theta)},\,
     \sin{\eta}\, \sin{\textstyle (\frac{2 \pi f x}{L}\!+\!\theta)},\,
     \cos{\eta} )  ,
\end{eqnarray}
and similarly for $\vec{\tilde{B}}$.
We have introduced $\theta$, so we can describe the 
 textures with a field component radial to the ring, i.e.\ $\theta=0$,
 as well as textures with a field component tangential to the ring,
 i.e.\ $\theta=\pi/2$.

\begin{figure}[H]
\centerline{\psfig{file=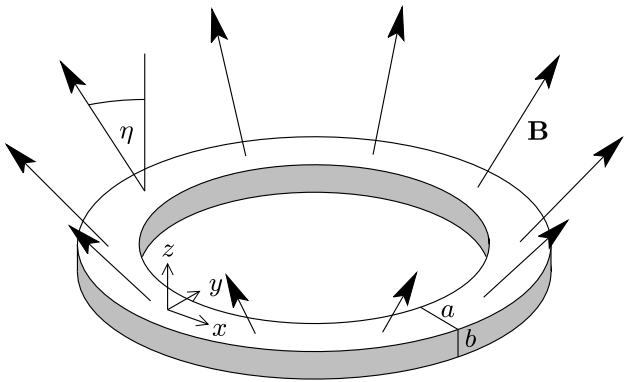,width=\figwidth}}
\caption[Ring with inhomogeneous field]{  
  A mesoscopic ring of width $a$ and height $b$
  in an inhomogeneous magnetic field
  with tilt angle $\eta$, winding once around the $z$-Axis.
 The texture of the magnetic field
  drawn here corresponds to Eq.~(\ref{eqnField})
  with $f=1$ and $\theta=0$.
}
\label{figRing}
\end{figure}

The starting point of our calculation is
 the conductance correlator derived in Ref.~\onlinecite{LSG}
 and given by
\begin{eqnarray}
\label{eqnUcfTrChi}
\dgUCF &  =   & \left(\frac{2 e^2 D}{h L^2}\right)^2  
  \int d\epsilon \, d\epsilon' n'(\epsilon) \, n'(\epsilon')
  \left\{
  \frac{1}{d}{\rm Tr}\, \hat{\chi}^C_\omega \hat{\chi}^{C\dagger}_\omega
\right. \nonumber \\  %noPP
& &  \left. %noPP
+2 \, {\rm Re}\,{\rm Tr}\, \hat{\chi}^C_\omega \hat{\chi}^{C}_\omega
 + \left[ \hat{\chi}^C \rightarrow \hat{\chi}^D \right]
   \right\} ,
\end{eqnarray}
where $n'(\epsilon)$ is the derivative of the Fermi function and
 $\hbar\omega=\epsilon-\epsilon'$.
The dimensionality of the system with respect to 
 the diffusive motion is denoted
 by $d$, which describes the relation of the mean free path $\ell$
 to the diffusion coefficient $D$, i.e.\ $D=\vfermi\ell/d$.
The propagators $\hat{\chi}^{C/D}$ can 
be evaluated explicitly by using the
operator equation Eq.~(\ref{diffEqnOmega}):
\begin{equation}
\label{eqnChi}
\hat{\chi}^{C/D}
= \frac{L^2}{(2\pi)^2 D} \:
   \frac{1}{i\tDeph + \gamma^{C/D} - h^{C/D}}
 .
\end{equation}
We have defined
$\tDeph = (L/2\pi L_T)^2 \, (\epsilon - \epsilon')/k T$,
with the thermal diffusion length $L_T = \sqrt{D\hbar\beta}.$\cite{lee}
The (non-hermitian) Hamiltonian is given by
\begin{equation}
\label{hamiltonian}
h^{C/D} = \frac{L^2}{(2\pi)^2} \frac{\partial^2}{\partial x^2}
  + i \kappa \vec{n} \cdot \bsigma_1
  - i \tilde{\kappa} \vec{\tilde{n}} \cdot \bsigma_2^{(*)}
,
\end{equation}
where the star means complex conjugation in $h^D$ and 
 where we have introduced an adiabaticity parameter~\cite{LSG,LSG99,langen}
\begin{equation}
\label{eqnKappa}
\kappa = \frac{\omegabohr}{D} \frac{L^2}{(2\pi)^2}
 ,
\end{equation}
and equivalently for $\tilde{\kappa}$ and $\omegabohrTilde$.
We have inserted a phenomenological damping constant 
$\gamma^{C/D} = (L/2\pi L_{C/D})^2$ expressed in terms of
 the magnetic dephasing length $L_{C/D}$\cite{aronov,LSG}:
\begin{equation}
\label{eqnLCD}
\gamma^{C/D} = 
  \frac{L^2}{(2\pi)^2L_\varphi^2} + \frac{1}{3(4\pi)^2} 
      \left(  \frac{A \, | B_z \pm \tilde{B}_z |}{2 \pi \phi_0} \right)^2
 .
\end{equation}
The first term of this damping constant incorporates the loss of phase due 
 to inelastic scattering events.
The second term takes into account
  magnetic flux penetration into the arms of the ring 
  with a finite width $a$ and a surface area $A=a L$,
  while the height $b$ is assumed to be small compared to $a$.
This field penetration leads to averaging over closed paths of 
 different lengths, each of which collects a different Aharonov-Bohm phase,
 resulting finally in dephasing.

Next we define the basis in which we evaluate the Hamiltonian $h^{C/D}$.
As done in Ref.~\onlinecite{langen} for the cooperon propagator, 
 we now introduce the operators
\begin{equation}
\label{eqnJCD}
 J^{C/D} = \frac{L}{2 \pi i}\frac{\partial}{\partial x} +
   \frac{1}{2} f (\sigma_{1z} \pm \sigma_{2z}),
\end{equation}
which commute with $h^{C/D}$.\cite{footnoteGeneralize}
 
We will now go to the basis of eigenvectors $\ket{j,\alpha\beta}_{C/D}$
 of $J^{C/D}$.
This basis is orthonormal with the following wave functions:
\begin{equation}
\label{eqnJOverlap}
\braket{x,\alpha'\beta'}{j,\alpha\beta}_{C/D} = 
  \frac{\delta_{\alpha'\alpha}\delta_{\beta'\beta}}{\sqrt{L}}\; 
  \exp\left\{ {\textstyle \frac{2\pi i x}{L}(j-\frac{f}{2}\alpha 
    \mp\frac{f}{2}\beta)   }   \right\}.
\end{equation}
Because of the periodic boundary conditions in $x$, 
the eigenvalues $j$ of $J^{C/D}$ have to  be integers.
The matrix elements of $h^{C/D}$ in the basis
$\{ \cdket{j,\uparrow\uparrow}, \cdket{j,\uparrow\downarrow},
 \cdket{j,\downarrow\uparrow},\cdket{j,\downarrow\downarrow} \}$
 become:
\begin{equation}
\label{eqnHMatrixElement}
  _{C/D}\!\braopket{j,\alpha\beta}{h^{C/D}}{j',\alpha'\beta'} {}_{C/D}
 =  \delta_{j j'}  \left( - h_j^{C/D} +  h_\sigma^{C/D} \right),
\end{equation}
where $h_j^C$ and $h_j^D$ are diagonal 4$\times$4 matrices with the entries
$\{(j-f)^2, j^2, j^2, (j+f)^2\}$, and 
$\{j^2, (j-f)^2, (j+f)^2, j^2 \}$, {\it resp.},
and the $\eta$, $\tilde{\eta}$ dependent matrices are

\ifpreprintsty\else\end{multicols}\emcsep\fi

\begin{equation}
\label{eqnHMatrix}
h_\sigma^{C/D}
 = 
\left(
\begin{array}{*{3}{c@{\!}}c}
i\kappa \,\cos{\eta } - i\tilde{\kappa }\,\cos{\tilde{\eta }} & -
   i\tilde{\kappa }e^{\mp i\theta}\,\sin{\tilde{\eta }} & 
   i\kappa e^{-i\theta} \,\sin{\eta } & 
   0 \\
-i\tilde{\kappa }e^{\pm i\theta}\,\sin{\tilde{\eta }} & 
   i\kappa \,\cos{\eta } + i\tilde{\kappa }\,\cos{\tilde{\eta }} &
   0 & 
   i\kappa e^{-i\theta}\,\sin{\eta } \\
i\kappa e^{i\theta} \,\sin{\eta } &
   0 & 
   -i\kappa \,\cos{\eta } - i\tilde{\kappa }\,\cos{\tilde{\eta }} &
   -i\tilde{\kappa }e^{\mp i\theta}\,\sin{\tilde{\eta }} \\
0 &
   i\kappa e^{i\theta}\,\sin{\eta } & 
   -i\tilde{\kappa }e^{\pm i\theta}\,\sin{\tilde{\eta }} & 
   -i\kappa \,\cos{\eta } + i\tilde{\kappa} \,\cos{\tilde{\eta }}
\end{array}
\right)
 .
\end{equation}

\ifpreprintsty\else\bmcsep\begin{multicols}{2}\fi

 To take the Aharonov-Bohm flux into account, we replace 
$j \rightarrow m = j - (\phi/\phi_0 \pm \tilde{\phi}/\phi_0)$, 
where $\phi, \tilde{\phi}$ are
the fluxes of the fields $\vec{B}, \vec{\tilde{B}}$ through the ring
and $\phi_0 = h/e$ is the magnetic flux quantum.\cite{fnPeriodicity}
Now it is straightforward to evaluate the exact conductance fluctuations
 $\dgUCF$
 by calculating the propagators by matrix inversion and inserting
 the result into Eq.~(\ref{eqnUcfTrChi}). 
This can be done with the help of the computer program Mathematica,
 which however 
 leads to lengthy expressions which
 we will not reproduce here.
We merely point out
 that the phase factors in $\theta$ cancel each other
 in $\dgUCF$ and $\dgMC$.

\subsection{Adiabatic Approximation}
\label{ssecUCFAd}
 To evaluate the adiabatic limit,
 we shall consider the regime of large magnetic fields
 with $B$ and $\tilde{B}$ of similar magnitude.
If we define $\Delta\kappa = \tilde{\kappa} - \kappa$,
 this adiabatic regime is described by
\begin{equation}
\label{eqnKappaBig}
  \kappa \gg 1 \quad \mbox{and} \quad  \kappa \gg |\Delta\kappa|.
\end{equation}
The exact propagators $\chi^{C/D}$ turn out to be rational functions
which are of order two in $\kappa$ in both numerator and denominator.
Now we will keep only the terms of highest order in $\kappa$;
 terms with large $j$ can be neglected as the sum over $j$ converges rapidly.
This leads us to the UCFs in the adiabatic regime:

\ifpreprintsty\else\end{multicols}\emcsep\fi

\begin{eqnarray}
\label{eqnUCFad}
\dgUCF_\adtag
& =  & \left(\frac{e^2}{h}\right)^2 \frac{1}{4 \pi^4}  
  \int d\epsilon\,d\epsilon' n'(\epsilon) \, n'(\epsilon')
  \sum_{j=-\infty}^{\infty}  \sum_{\alpha=\pm 1}
  \left( G^\adtag_{\alpha,C} + G^\adtag_{\alpha,D} \right)
\\
\nonumber
G^\adtag_{\alpha, C/D} & = &
\frac{1}{d} \left\{ {{\left( \tDeph  - \alpha  \Delta \kappa  \right) }^2} + 
          \deltaCDSq{-} + P  \right\} 
\times  \left\{  
 \left[ {{\left( \tDeph  - \alpha  \Delta \kappa \right) }^2} + 
               \deltaCDSq{-} - P  \right] 
      \right.
\\
& & 
  \qquad \left. \cdot  \left[ {{\left( \tDeph + \alpha  \Delta \kappa  \right) }^2}
           +  \deltaCDSq{} - P  \right] 
     + 4 P  \left[ {\tDeph^2} + f^2 
m^2 \left( \cos{\eta } \pm \cos{\tilde{\eta }} \right)^2  \right] \right\}^{-1}
\nonumber \\
& + &
 2 {\rm Re} \bigglb[
\left\{ \left[ i \tDeph   - i \alpha  \Delta\kappa 
           + \deltaCD{-} \right]^2  + P \right\}
\nonumber \\
\label{eqnGad} & &  
  \times  \left\{  \left[  i \tDeph  - i \alpha  \Delta\kappa 
      + \deltaCD{-} \right] 
   \left[  i \tDeph   + i \alpha  \Delta\kappa 
     + \deltaCD{} \right] - P \right\}^{-2} \biggrb]
,
\end{eqnarray}
\ifpreprintsty\else \clearpage \begin{multicols}{2}\noindent \fi  
where

\begin{eqnarray}
\label{eqnP}
P & = &
   \frac{f^4}{4} \,\sin^2{\eta } \,\sin^2{\tilde{\eta}}
,
\\
\label{eqnDeltaj}
 \deltaCD{} & = &
   \gammatilde
  + \left( m - \frac{f}{2} \, \alpha  \,\cos{\eta } \mp
              \frac{f}{2} \, \alpha  \,\cos{\tilde{\eta}} \right)^2
  ,
\end{eqnarray}
with

\begin{eqnarray}
\label{eqnTildeGamma}
 \gammatilde &=& \gamma^{C/D}
  +\frac{f^2}{4} \,\sin^2{\eta} + \frac{f^2}{4} \,\sin^2{\tilde{\eta}}
  .
\end{eqnarray}

The sum over $\alpha$ has been introduced here artificially to 
 facilitate the following interpretation.
As it is also seen in 
 Ref.~\onlinecite{LSG99} for the case of the magnetoconductance $\dgMC$,
 the terms $f^2(\sin^2{\eta} + \sin^2{\tilde{\eta}})/4$ 
 in Eq.~(\ref{eqnTildeGamma}) act as additional dephasing sources
 and are here absorbed in 
 the phenomenological dephasing parameter 
 $\gammatilde$.
However, in Eq.~(\ref{eqnGad}) there are further 
 $\eta$, $\tilde{\eta}$-dependent terms $P$,
 which cannot be formally absorbed in $\gammatilde$.
$P$ reduces the effect of the additional dephasing terms 
 in Eq.~(\ref{eqnTildeGamma}),
 as we can see by the following numerical evaluation.
We consider equal fields $\vec{B}=\vec{\tilde{B}}$ and low temperatures,
 thus $\Delta\kappa,\,\omega=0$,
 and assume $\eta$, $\tilde{\eta}$ to be close to $\pi/2$.
Then we estimate the amplitude of the Aharonov-Bohm oscillations
 by taking the difference between
 the values of $G^\adtag_{\alpha,C}$ [Eq.~(\ref{eqnGad})]
 for the two phases $m=0$ and $m=\pm 1/2$
 (i.e.\ we are considering only the main contributions in the 
  sum over~$j$ [Eq.~(\ref{eqnUCFad})]).
We then see by numerical evaluation
 that the oscillations are suppressed if we set $P=0$ instead of 
 using Eq.~(\ref{eqnP}),
 thus $P$ indeed reduces dephasing.

We can compare now with previous calculations\cite{LSG}
 where the UCFs $\dgUCF_\lsgtag$ have been derived for 
 arbitrary textures and adiabatic evolution of the spin.
These results can be recovered from Eq.~(\ref{eqnUCFad})
 by the replacement
 $\gammatilde \to \gammalsg$ and $P\to 0$.
The dephasing terms due to inhomogeneous fields
 coupling to the spin
 [see Eqs.~(\ref{eqnP}), (\ref{eqnTildeGamma})] were not explicitly given
 in Ref.~\onlinecite{LSG};
 to account for such dephasing 
 these terms must be included in 
 the phenomenological parameter $\gammalsg$,
 and thus $\gammalsg\neq\gamma^{C/D}$ and $\gammalsg\neq\gammatilde$
 in general.\cite{LSG99}

We also recognize a strong simplification in the special case
 where one field is homogeneous, $\eta=0$, i.e., $P$ vanishes.
Thus the comparison of $\dgUCF_\adtag$ with 
 the solution for arbitrary textures $\dgUCF_\lsgtag$ 
 yields the simple relation $\gammalsg = \gammatildez$.
Finally we note that in this case
 the dephasing due to the orientational inhomogenity of 
 $\vec{\tilde{B}}$ measured by the winding $f$
 grows like $f^2 \,\sin^2{\tilde{\eta}}$ [cf. Eq.~(\ref{eqnTildeGamma})].

\subsection{Finite Temperatures}
\label{ssecFiniteTemperatures}

Now we consider the effects of finite temperatures $T\!>\!0$
 on the UCFs $\dgUCF_\adtag$ in the adiabatic regime.
In the case of $\eta=0$, i.e.\ $P=0$, the factors containing $\deltaCD{-}$ 
 in Eq.~(\ref{eqnGad}) cancel, so we obtain
\begin{eqnarray}
\label{eqnGadPZero}
\left. G^\adtag_{\alpha, C/D} \right| _{\eta=0}  &=& 
\frac{1}{d}  \left\{  
  {{\left( \tDeph + \alpha  \Delta \kappa  \right) }^2}
           +  \deltaCDSq{} \right\}^{-1}
\nonumber \\ & &  %noPP
 +  2 {\rm Re}
\left\{   i \tDeph   + i \alpha  \Delta\kappa 
     + \deltaCD{} \right\}^{-2}
 .
\end{eqnarray}
This strong simplification allows us to evaluate
the integrals over $\epsilon$ and $\epsilon'$ in Eq.~(\ref{eqnUCFad})
explicitly by using standard Matsubara techniques,
  as described in App. \ref{appMatsubara},
 and we obtain for the UCFs 
 $\dgUCF_\adtag  = \dgUCF_{\adtag,\,C} + \dgUCF_{\adtag,\,D}$,  
\begin{eqnarray}
\label{eqnUCFTfinite}
\nonumber
\left. \dgUCF_{\adtag,\, C/D}
  \right|_{\eta=0} 
& & =  \!
 \left(\frac{e^2}{h}\right)^2 \!\! \frac{1}{8\pi^6} \left(\frac{L^2}{L_T^2}\right)^2 {\rm Re}  \sum_{\alpha=\pm1} \left. \sum_{j,n,m}\right.'
\\
\nonumber
& & \!\!\!\!\!\!\!\!
  \left\{ 
   \frac{1}
    {d \, \delta_{\alpha}^{C/D}(j) \cdot 
   \left[ \frac{L^2}{4\pi L_T^2}(m\!+\!n)\! +\! \delta_{\alpha}^{C/D}(j) \! -\! i \alpha\Delta\kappa  \right]^3 }
\right. \\ & &  \left. %noPP
   +  \frac{6}
  {\left[ \frac{L^2}{4\pi L_T^2}(m\!+\!n)\! +\! \delta_{\alpha}^{C/D}(j) \! -\! i \alpha\Delta\kappa  \right]^4}
  \right\}
  .
\end{eqnarray}
Here $n$ and $m$ are odd, positive integers.
For plotting, it is advantageous to calculate the sum 
 in Eq.~(\ref{eqnUCFTfinite})
 analytically, which gives an expression containing Psi-functions.

We can now obtain a qualitative criterion when the thermal dephasing
 effects can be ignored.
If we ignore thermal effects, i.e.\ assume low temperatures,
 we can simplify our calculation leading to Eq.~(\ref{eqnUCFTfinite})
 by replacing  $n'(\epsilon)$ by a 
 delta function $\delta(\epsilon)$ in Eq.~(\ref{eqnUCFad}).
This yields for $\eta=0$ the same result as applying
 Poisson's summation formula
to Eq.~(\ref{eqnUCFTfinite}) in order 
 to replace the summations over $n$ and $m$ by integrations.
We are only allowed to perform this step if the summand varies
slowly in $n$,~$m$, which is the case for  $L_T^2 \gg 2\pi L_{C/D}^{2}$.
 From a physical point of view, this is an evident requirement:
 the smearing of the conductance fluctuations due to nonzero temperatures,
 described by the thermal diffusion length $L_T$,
 can only be neglected if the dephasing lengths 
 related to inelastic scattering or penetrating magnetic fields 
 are much shorter than $L_T$.

In App.~\ref{secUcfLimits} 
 we evaluate the dephasing behavior of the UCFs $\dgUCF_\homtag$
 for homogeneous fields and finite temperatures.
Then we confirm the result of  Ref.~\onlinecite{LSG}
 [Eq.~(\ref{eqnUcfTrChi})]
 and show that our calculation in the homogeneous limit indeed reproduces
 known results.\cite{lee,altshuler,beenakker}

\section{Berry phase and Adiabaticity}
\label{secBerryAd}

\subsection{Magic Angles---Qualitative Criterion 
            for Adiabaticity}
\label{ssecQual}

We now consider the qualitative effects of the Berry phase on the conductance
fluctuations $\dgUCF$.
They emerge from the Berry phase in $\deltaCD{}$ in the adiabatic
  solution [Eq.~(\ref{eqnUCFad})]
 and lead to vanishing Aharonov-Bohm oscillations at special ``magic'' tilt
 angles of the magnetic fields.
This effect has some similarities with the phenomenon of beating, 
 where the superposition of 
 two oscillations with different but fixed frequencies leads to a periodic
 vanishing of the envelope.
However, in our case we have two frequencies which will 
 change when the perpendicular field $B_z$
 is increased, since then the Berry phase is altered, too.
Thus a suppression of the Aharonov-Bohm oscillations can only be observed at
 two special tilt angles of the magnetic field, 
 i.e.\ the Berry phase has a highly non-periodic effect on the
 envelope of these oscillations as a function of $B_z$.

 From now on we shall only study the experimentally realizable field texture
 with one winding, $f=1$.
 The other configurations with $f>1$ are solely of academic
 interest.
 To illustrate expected experimental results, 
 we will use some material
 parameters recently determined.\cite{mohanty}
The 
 sample Au-1 given in Table I of Ref.~\onlinecite{mohanty} 
 has the values $D=9\times10^{-3}\:{\rm m}^2\,{\rm s}^{-1}$
 and $L_\varphi = \sqrt{D \tau_\varphi} = 5.54 \:\mu{\rm m}$.
We assume a ring with diameter of $4 \:\mu {\rm m}$, 
 so $L=12.6 \:\mu{\rm m}$, and an arm width $a = 60 \,{\rm nm}$,
 which lies well within present-day experimental reach. 
Finally we assume low temperatures, i.e.\ $L_T \gg L,L_\varphi$, 
 so we can ignore the dephasing due to thermal fluctuations.

Now we shall consider two equal fields, so no phase terms appear in
 the diffuson contribution $\dgUCF_D$.
The cooperon contribution $\dgUCF_C$ is $h/2e$ periodic in the 
 magnetic flux, as a shift of 
$m=\phi/\phi_0+\tilde{\phi}/\phi_0+j=2\phi/\phi_0+j$
by 1 is absorbed in the sum over $j$ in Eq.~(\ref{eqnUCFad}).
For the next argument we take the dephasing due to inhomogeneous
 fields only phenomenologically into account, 
 i.e.\ we use the result $\dgUCF_\lsgtag$ from Ref.~\onlinecite{LSG}
 or equivalently set $P=0$ [Eqs.~(\ref{eqnUCFad}, \ref{eqnGadPZero})],
 so the factors containing $\deltaCD{-}$ cancel in Eq.~(\ref{eqnGad}).
If the tilt angle $\eta$ is such that $\cos{\eta}=1/4$,
 the phase dependent term
 in $\deltaCD{}$ [Eq.~(\ref{eqnDeltaj})]
 becomes $m - \alpha/4$. 
One sees that in this special case shifting $m$ by $1/2$ does not affect the 
 value of $\dgUCF_C$, 
 as it leads solely to an exchange $\alpha \to -\alpha$.
The very same argument applies to $\cos{\eta}=3/4$.
Thus, for these {\it magic angles} $\eta$, where $\cos{\eta}=1/4,3/4$,
 the UCFs $\dgUCF$ are $h/4e$ periodic and 
 therefore their 
 {\it power spectrum shows a vanishing $h/2e$ amplitude.}
If we take the exact solution in the adiabatic regime $\dgUCF_\adtag$ 
 instead of $\dgUCF_\lsgtag$,
 the magic angles are still present, but at shifted values.
The angle at $\cos{\eta}=3/4$ is nearly unaffected,
 as $P\approx0.05$ is very small at this angle.
The suppression of the Aharonov-Bohm oscillations
 is illustrated in Fig.~\ref{figh2e}
 (see also Sec.~\ref{ssecPSTechniques} and Fig.~\ref{figUcfH2eOsc})
 by plotting the $h/2e$ amplitude of the exact solution $\dgUCF$
 with varying tilt angle $\eta$ 
 and for different radial field components.
As one can readily see from Fig.~\ref{figh2e},
 the effect described here is fully developed for
 $B\geq 200\:{\rm G}$. 
For smaller fields, the $h/2e$ amplitude
  does not completely vanish at the magic angles, as adiabaticity is
  not yet reached.
It should be noted that even if the adiabatic regime is not fully 
 reached, an effect of the Berry phase is still visible
 as a distinct non-monotonic behavior of the UCFs $\dgUCF$ as 
 a function of the tilt angle $\eta$,
 unlike the UCFs for a configuration with a homogeneous field texture 
 (also shown in Fig.~\ref{figh2e}).

\begin{figure}[H]
\centerline{\psfig{file=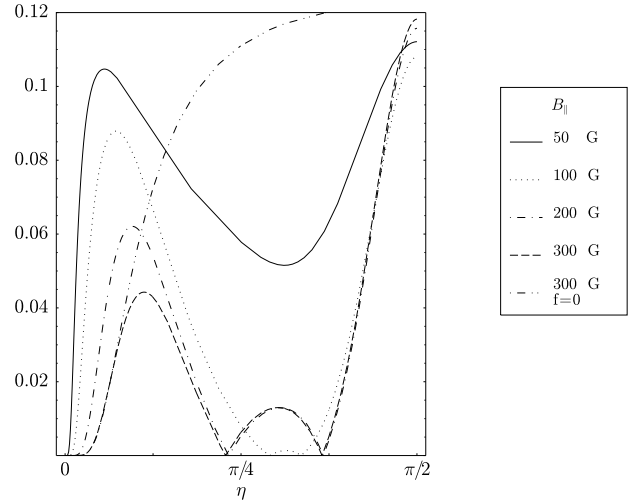,width=\figwidth}}
\caption[$h/2e$ oscillations of the UCFs]{  
The normalized amplitudes of the $h/2e$ oscillations 
 in the UCFs $\dgUCF$,
 as a function of the tilt angle $\eta$.
The magnetic fields are chosen equal, i.e.\ $\vec{B}=\vec{\tilde{B}}$,
 and wind once around the ring (i.e.\ $f=1$).
The power spectrum of the exact UCFs $\dgUCF$ has been calculated at every tilt
 angle $\eta$  by varying  the Aharonov-Bohm flux 
 $0\leq \phi=\tilde{\phi} \leq 1$.
The component of the $h/2e$ oscillation in this spectrum was then normalized 
 by the 0th order Fourier component and is plotted here
 as a function of $\eta$.
Four configurations of radial fields $B_{||}=\tilde{B}_{||}$ are
 shown; the perpendicular field components $B_z=\tilde{B}_z$ are
 determined by the tilt angles $\eta=\tilde{\eta}$.
These field components and so 
 also $\gamma^C$, as it depends on the arm-penetrating field,
 increase for small $\eta$.
The strong dephasing $\gamma^C$ 
 at $\eta \approx 0$ can be observed as vanishing oscillations.
The most remarkable effects show up for the stronger fields
 $B_{||} = 200\:{\rm G},\, 300\:{\rm G}$ 
 at the magic angles $\eta=0.72, \, 1.15$.
Here the Berry phase eliminates the
 $h/2e$ oscillations, as it is described in Section~\ref{ssecQual}.
For comparison, we also show the conductance fluctuations for a
 homogeneous field, i.e.~setting $f=0$.
We here set $T=0$
 and used the material parameters 
 $L=12.6 \:\mu {\rm m}$, 
 $a = 60 \,{\rm nm}$,
 $D=9\times10^{-3}\:{\rm m}^2\,{\rm s}^{-1}$, and
 $L_\varphi =  5.54 \:\mu{\rm m}$.
}
\label{figh2e}
\end{figure}

\ifpreprintsty \else \clearpage \fi %noPP

\begin{figure}[H]
\centerline{\psfig{file=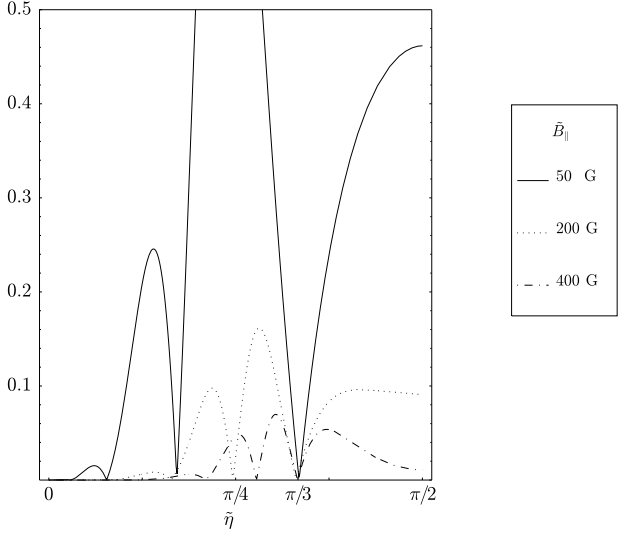,width=\figwidth}}
\caption{
The normalized amplitudes 
 of the $h/e$ oscillations in the UCFs $\dgUCF$,
 with $\eta=0$, as a function of the tilt angle $\tilde{\eta}$.
The field were taken as 
 $\vec{B}=(2/\sqrt{3})\tilde{B}_\parallel\,\vec{e}_z$,
 $\tilde{B}_\parallel = {\rm const.}$,
and $\tilde{B}_z$ was determined through the tilt angle $\tilde{\eta}$.
We use the same methods and parameters 
 as described in Fig.~\ref{figh2e}
 for $\tilde{B}_\parallel = 50\:{\rm G}$, $200\:{\rm G}$, and $400\:{\rm G}$.
We notice that the $h/e$ oscillations become suppressed by the Berry phase
 at the magic angle $\cos{\tilde{\eta}}=\pi/3$.
}
\label{figDiffHe}
\end{figure}

Another interesting situation arises for $B \neq \tilde{B}$.
Now, phase effects from the diffuson contribution to $\dgUCF$
 emerge and remain present even for large fields,
 since the dephasing due to flux penetrating the arms of the ring
 depends only on the difference of the fields
 and not on the sum as for the cooperon contribution,
 see Eq.~(\ref{eqnLCD}).
For illustration, we consider the configuration where $\vec{B}$ 
  is homogeneous with $\eta=0$.
The other field $\vec{\tilde{B}}$ is assumed to 
 have a radial component so that for a
 tilt angle $\tilde{\eta}=\pi/3$ the magnitudes of both fields are equal,
 i.e.\ $\tilde{B}_{||} = (\sqrt{3}/2) B_z$.
In the adiabatic approximation $\dgUCF_\adtag$ [Eq. (\ref{eqnUCFad})] 
 $P$ vanishes, 
  yielding the simple relation Eq.~(\ref{eqnTildeGamma})
  between the dephasing due to the inhomogeneous
  field textures and $\gamma^{C/D}$:
  the effective dephasing will be increased by $3/16$ at the most interesting
  angle, $\tilde{\eta}=\pi/3$, in the situation considered here.
 The contribution of the penetrating fields to $\gamma^{C/D}$ 
 will be three times larger for the cooperon than for the diffuson, 
 as can be seen from Eq.~(\ref{eqnLCD}).
Varying $\tilde{B}_z$ changes
 the Aharonov-Bohm phase $\tilde{\phi}/\phi_0$,
 while $\phi/\phi_0={\rm const.}$,
 leading to $h/e$ oscillations.
At $\tilde{B}_z = B_z/2$ two features are worth mentioning.
First, the magnitudes of both fields become equal, therefore
 $\Delta\kappa$ vanishes and so the second part of
 the criterion in Eq.~(\ref{eqnKappaBig}) is fulfilled
 and we can use the adiabatic approximation 
 $\dgUCF_\adtag$ [Eq.~(\ref{eqnUCFad})].  
Second, we have $\cos{\tilde{\eta}}=1/2$, 
 so the phase dependent
 terms $m\mp\alpha/4$ arise in $\deltaCD{}$,
 as can be seen from Eq.~(\ref{eqnDeltaj}).
With the same argument as above, the
 UCFs $\dgUCF$ become $h/2e$ periodic at this magic angle $\pi/3$, so the
 $h/e$ amplitude vanishes in the power spectrum.
We note that, in the adiabatic regime, this magic angle is exact,
 since for the configuration $\eta=0$ we have $\dgUCF_\adtag = \dgUCF_\lsgtag$.
This is shown in Fig.~\ref{figDiffHe}, again as a function of
 the tilt angle $\tilde{\eta} = \cot{ (\tilde{B}_z / B_z )}$,
 see also Sec.~\ref{ssecPSTechniques} and Fig.~\ref{figUcfHeOsc}.

\subsection{Quantitative Criterion for Adiabaticity}
\label{ssecQuant}

In order to obtain a quantitative criterion for adiabaticity,
we  numerically compare the exact solution of 
 the conductance fluctuations $\dgUCF$
 with the adiabatic approximation $\dgUCF_\adtag$ [Eq.~(\ref{eqnUCFad})].
We take equal magnitudes for both fields, i.e.\ $B=\tilde{B}$.
We search for a minimal $\kappa_{\rm min}$  so that the relative 
 difference $\big| \dgUCF - \dgUCF_\adtag \big| \big/ \dgUCF$
 is below a certain value.
This is done with a bisection algorithm (in $\kappa$) and
 by sampling over the parameter subspace
$[0,\pi/2]^2 \times [0,1]^2 \times [\frac{1}{100}, 10]^2 
 \subset \{ (\eta,\, \tilde{\eta},\,
  \phi/\phi_0,\, \tilde{\phi}/\phi_0,\,
  \gamma^C,\, \gamma^D )\}$
with a grid resolution of 10 intersections in the first four dimensions.
A finer resolution has been chosen for $\gamma^{C/D}$.
As can be seen from Fig.~\ref{figKmin}, 
 for $0.01 \leq \gamma^D \leq 1,\, \gamma^D\leq\gamma^C$ 
 and a field strength such that $\kappa \geq 3$,
 the numerical values for $\dgUCF$ and $\dgUCF_\adtag$
 are already within five percent of each other.

\begin{figure}[H]
\centerline{\psfig{file=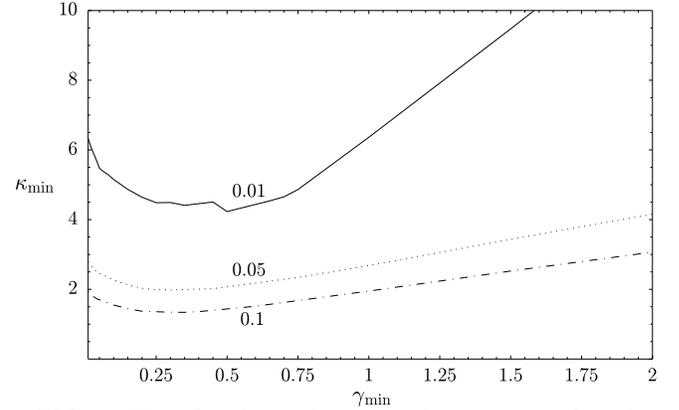,width=\figwidth}}
\caption{This plot shows the minimal $\kappa_{\rm min}$ required
 so that the normalized
 difference $\big| \dgUCF - \dgUCF_\adtag \big| \big/ \dgUCF$
 is smaller than $0.01$, $0.05$, and $0.5$;
 i.e.\ the plot shows for which magnitudes of the magnetic field the
 exact solution of the UCFs $\dgUCF$ agrees with the 
 adiabatic approximation $\dgUCF_\adtag$ [Eq.~(\ref{eqnUCFad})] 
 to a certain accuracy.
$\kappa_{\rm min}$ is plotted against 
 $\gamma_{\rm min} = {\rm min}\{\gamma^C,\,\gamma^D\}$; 
 as the two fields $\vec{B},\,\vec{\tilde{B}}$ may have different orientations,
 $\gamma^D$ can become larger than $\gamma^C$.
As $\dgUCF$ vanishes for large $\gamma^{C/D}$, 
 our normalization is no longer 
 well defined for $\gamma^{C/D} \protect\gtrsim 1$
 and the value for $\kappa_{\rm min}$ diverges. }
\label{figKmin}
\end{figure}

However, as we are interested in the 
 Aharonov-Bohm oscillations rather
 than in the absolute value of the UCFs $\dgUCF$, 
 we now use a different method of comparison: 
We consider the oscillations  in the conductance fluctuations
 resulting from different Aharonov-Bohm fluxes through the ring.
As a measure for accuracy we take the relative error of these amplitudes,
 i.e.\ 
\begin{eqnarray}
\label{eqnDeltaKmin}
& & \Delta(\kappa, \gamma^C, \gamma^D, \eta, \tilde{\eta})
\\ \nonumber &   %noPP
 = 
&  %noPP 
 \frac{ {\displaystyle \max_{\phi, \tilde{\phi}} } \left|
   (\dgUCF         - \dgUCF          |_{\phi=\tilde{\phi}=0} )
  -(\dgUCF_\adtag- \dgUCF_\adtag |_{\phi=\tilde{\phi}=0} )
  \right| }{
   {\displaystyle \max_{\phi, \tilde{\phi}} } \left|
\dgUCF - \dgUCF |_{\phi=\tilde{\phi}=0} \right|
  _{\eta,\tilde{\eta} = 0}
}.
\end{eqnarray}
Again we search for a minimal $\kappa_{\rm min}$ so that $\Delta$ is bounded 
 from above by
 a certain percentage over the whole parameter subspace.
We notice from the results shown in Fig.~\ref{figKminEta}
 that in the regime with only moderate damping $\gamma^C=\gamma^D=0.1$,
 adiabaticity is already reached at $\kappa \sim 2$.
If we put this in the context of the experimental parameters given in the
 beginning of Sec.~\ref{ssecQual}, we expect adiabaticity to 
 be fully reached at magnetic fields
 of magnitude larger than $500\:{\rm G}$.
By comparing this value with Fig.~\ref{figh2e}, we note that the
 qualitative effect of the Berry phase can already be seen
 for fields which are an order of a magnitude smaller,
 i.e.\ for $B,\,\tilde{B} \gtrsim 50\:{\rm G}$.

\begin{figure}[H]
\centerline{\psfig{file=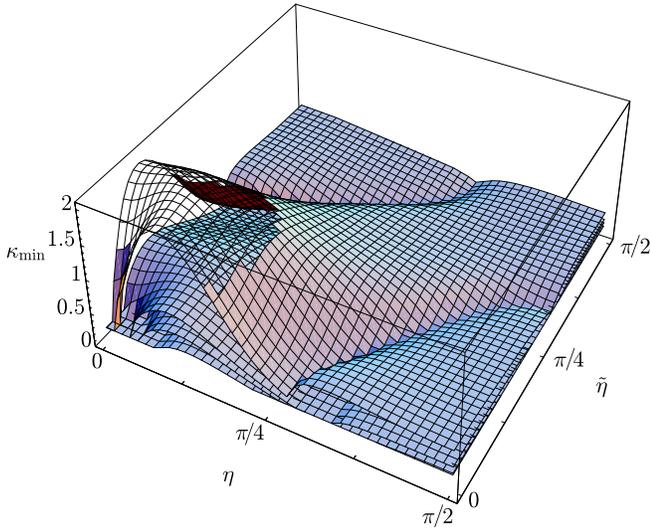,width=\figwidth}}
\caption{ Here the quality of the 
 adiabatic approximation $\dgUCF_\adtag$ [Eq.~(\ref{eqnUCFad})] 
 in describing the Aharonov-Bohm oscillations is shown.
We used Eq.~(\ref{eqnDeltaKmin}) and set $\gamma^C=\gamma^D=0.1$.
 The surfaces shown are, from top to bottom,
  the minimal value of $\kappa_{\rm min}$ required for an agreement 
  $\Delta < 0.01$, $0.05$, $0.1$, and $0.2$ [Eq.~(\ref{eqnDeltaKmin})].
 As expected, for $\eta=\tilde{\eta}=0$ we have $\dgUCF = \dgUCF_\adtag$.
 For tilt angles $\eta,\,\tilde{\eta} \approx \pi/2$, 
  the agreement is obtained at low $\kappa_{\rm min}$,
  whereas at $\eta=\tilde{\eta}\approx0.25$ larger fields are
  necessary.
}
\label{figKminEta}
\end{figure}

We now discuss the effects of different parameters on $\kappa$ 
 and on the minimal magnetic fields required to reach adiabaticity,
 thus indicating favorable experimental setups.
If we consider rings of increasing circumference $L$, we can see
 from Eq.~(\ref{eqnKappa})
 that the minimal magnetic field strength needed decreases
 as $B_\adtag \propto L^{-2}$.
However, to observe the Berry phase, dephasing must not be too strong,
 so the condition $L \lesssim 2L_{C/D}$ 
 should still be met.
We note that for two equal fields, 
 the first term of $\gamma^C \propto L_C^{-2}$ 
 in Eq.~(\ref{eqnLCD}) depends on $L^2$,
  which restrains us from taking $L> 2L_\varphi$,
 whereas the second one depends for $B=B_\adtag$ on $L^{-2}$.
So not only the high magnetic fields needed for adiabaticity, 
 but also the small arm widths $a$ required to minimize
 strong dephasing due to the penetrating flux, disfavors experimental setups
 with very small $L$.

Introducing more impurities and thus decreasing the diffusion coefficient $D$
 leads to slower motion of the electrons around the ring, giving
 their spins more time to adjust to the local magnetic texture.
 Thus, the field strengths required for adiabaticity to occur decrease
 as $B_\adtag\propto D$,
 which can be seen from Eq.~(\ref{eqnKappa}).
However, such slow diffusion also leads to shorter dephasing lengths 
 $L_T,\,L_\varphi \propto D^{1/2}$; 
 assuming that $\tau_\varphi$ remains constant.
 To avoid such an additional dephasing, i.e.\ leaving $\gamma^{C/D}$ unaffected,
 the sample size must also be decreased as $L \propto D^{1/2}$.
Thus, because of $\kappa \propto D^{-1}L^2$, no net decrease of the required 
 fields for adiabaticity can be gained by decreasing the diffusion coefficient.

\section{Exact calculations with Spin-Orbit Interaction in Diffusive Limit}
\label{secSOExact}

We turn now to the discussion of Berry phases induced 
 by spin-orbit interaction.
Instead of considering an inhomogeneous field,
 we use here an effective (non-hermitian) Hamiltonian
\begin{eqnarray}
\label{eqnHSO}
 h^{C/D}_{\SO} &=& \frac{L}{(2\pi)^2} \frac{\partial^2}{\partial x^2}
  + i\kappa \sigma_{1z} - i\tilde{\kappa}\sigma_{2z}
\nonumber \\ & &  %noPP
 + i \frac{\alpha}{\hbar^2} \frac{L^2}{D (2\pi)^2}
       (\vec{e}_z \times \bsigma^{(*)})\cdot \vec{p}
  ,
\end{eqnarray}
 with spin-orbit interaction,
 using a coupling constant $\alpha$ as defined in Ref.~\onlinecite{rashba},
 and with a Zeeman term from an external magnetic field,
 which is perpendicular to the ring plane. 
One arrives at this Hamiltonian 
 by starting from the 
 Feynman path integral representation of the transition amplitude 
 with spin-orbit coupling, as it is given in Ref.~\onlinecite{Schmid}.
One can then formally decouple orbital and spin motion, and following
 the steps given in App.~A of Ref.~\onlinecite{LSG},
 one arrives at the effective Schr\"odinger equation for 
 the cooperon propagator with the Hamiltonian $h^C_\SO$.
The equation with $h^D_\SO$ for the diffuson, 
 which will be required in Sec.~\ref{ssecSOUCF},
 can be obtained by applying the techniques explained in App.~\ref{secCdDiffEqn}.

Note that in Eq.~(\ref{eqnHSO}) the momentum operator is still in
 the Cartesian coordinate system.
Now we adopt a polar coordinate system, with
 $(x',\,y')=(r\,\cosx{} ,\, r\,\sinx{})$ and
 $ ( \partial_{x'},\, \partial_{y'})
  = (-\frac{1}{2}\{\sinx{},\,\partial_x \},\: 
     \frac{1}{2} \{ \cosx{},\,\partial_x \})$,
 where $x$ denotes the position along the ring and runs from $0$ to $L$.
The curly braces denote the anticommutator,
 which ensures the hermiticity of the momentum operator.
We now have

\begin{eqnarray}
h_\SO^{C/D} &=&
  \frac{L^2}{(2\pi)^2} \frac{\partial^2}{\partial x^2}
  + i \kappa \sigma_{1z} - i \tilde{\kappa} \sigma_{2z}
\nonumber \\
 &&+\frac{\alpha}{\hbar} \frac{L^2}{D (2\pi)^2} 
    \left. \frac{1}{2} \right\{
      \sigma_{1x} \cosx{} 
     +\sigma_{1y} \sinx{} 
   \nonumber \\ && \quad %noPP
    \left.
     -\sigma_{2x} \cosx{} 
   \mp\sigma_{2y} \sinx{} 
  ,\: \frac{\partial}{\partial x}  \right\}
.
\end{eqnarray}
 To diagonalize the Hamiltonian, we follow the ideas used above
 and use the operators defined in 
 Eq.~(\ref{eqnJCD}), but now with $f=\tilde{f}=1$:
\begin{equation}
 J^{C/D} :=
  \frac{L}{2 \pi i} \frac{\partial}{\partial x}+
   \frac{1}{2} \sigma_{1z} \pm  \frac{1}{2} \sigma_{2z}
    ,
\end{equation}
which commute with the Hamiltonians $h_\SO^{C/D}$,
as can be seen using 
 $\left[\{ n(x),\, \partial_x\} ,\,\partial_x\right]
   = -\{n'(x),\,\partial_x\}$.
We can now calculate the matrix elements of $h_\SO^{C/D}$ in the
 basis defined in Eq.~(\ref{eqnJOverlap}), with  $f=\tilde{f}=1$,
 as

\ifpreprintsty\else\end{multicols}\emcsep\fi

\begin{equation}
\label{eqnHSOC}
\braopket{j,\alpha\beta}{h_\SO^C}{j',\alpha'\beta'}
 = 
\delta_{jj'}
\left(
\begin{array}{*{3}{c@{\:}}c}
-(j-1)^2 +i\kappa\!-\!i\tilde{\kappa} & %++,++
   iS\left(j-\half\right) & %++,+-
   iS\left(-j+\half\right)  & %++,-+
   \:\:\:\:\:0\:\:\:\:\: \\ %++,--
iS\left(j - \half\right)  & %+-,++
   -j^2 +i\kappa+i\tilde{\kappa}& %+-,+-
   \:\:\:\:\:0\:\:\:\:\: & %+-,-+
   iS\left(-j-\half\right) \\ %+-,--
iS\left(-j+\half\right) & %-+,++
   \:\:\:\:\:0\:\:\:\:\: & %-+,+-
   -j^2 -i\kappa-i\tilde{\kappa}& %-+,-+
   iS\left(j +\half\right) \\ %-+,--
\:\:\:\:\:0\:\:\:\:\: & %--,++
   iS\left(-j-\half\right)  & %--,+-
   iS\left(j +\half\right)   & %--,-+
   -(j+1)^2 -i\kappa\!+\!i\tilde{\kappa} %--,--
\end{array}
\right)
 ,
\end{equation}
and

\begin{equation}
\label{eqnHSOD}
\braopket{j,\alpha\beta}{h_\SO^D}{j',\alpha'\beta'}
 = 
\delta_{jj'}
\left(
\begin{array}{*{3}{c@{\:}}c}
-j^2 +i\kappa\!-\!i\tilde{\kappa} & %++,++
   iS\left(j-\half\right) & %++,+-
   iS\left(-j-\half\right)  & %++,-+
   \:\:\:\:\:0\:\:\:\:\: \\ %++,--
iS\left(j - \half\right)  & %+-,++
   -(j\!-\!1)^2  +i\kappa\!+\!i\tilde{\kappa}& %+-,+-
   \:\:\:\:\:0\:\:\:\:\: & %+-,-+
   iS\left(-j+\half\right) \\ %+-,--
iS\left(-j-\half\right) & %-+,++
   \:\:\:\:\:0\:\:\:\:\: & %-+,+-
   -(j\!+\!1)^2  -i\kappa\!-\!i\tilde{\kappa} & %-+,-+
   iS\left(j +\half\right) \\ %-+,--
\:\:\:\:\:0\:\:\:\:\: & %--,++
   iS\left(-j+\half\right)  & %--,+-
   iS\left(j +\half\right)   & %--,-+
   -j^2 -i\kappa\!+\!i\tilde{\kappa} %--,--
\end{array}
\right)
 .
\end{equation}

\ifpreprintsty\else\bmcsep\begin{multicols}{2}\fi

In Eqs.~(\ref{eqnHSOC}) and (\ref{eqnHSOD}), we have introduced a
 dimensionless spin-orbit coupling parameter
\begin{equation}
\label{eqnSOCoupling}
  S = \frac{\alpha}{\hbar D} \frac{L}{2\pi}.
\end{equation}

By comparing Eqs.~(\ref{eqnKappa}) and (\ref{eqnSOCoupling}),
 we note that while $\kappa$ is quadratic in $L$,
 the parameter $S$ is only linearly dependent on $L$.
If we define an effective field angle for diffusive motion with
 spin-orbit coupling

\begin{equation}
 \tan{\eta_{\SO}} = S/\kappa
  ,
\end{equation}  
and anticipate the Berry phase to be of the form $\Phi^g = \cos{\eta}$,
we obtain for $S\gg\kappa$ the dependency 
 $\Phi^g \approx \kappa/S \propto L$. 
Thus the phase can now be enhanced by increasing the size of the ring.
However, the phase cannot be increased arbitrarily; for large $L$,
 the assumption $S\gg\kappa$ becomes invalid.

\subsection{Magnetoconductance}
\label{ssecSOMC}

We shall now calculate the magnetoconductance
 with the formula from Ref.~\onlinecite{LSG}
\begin{equation}
 \dgMC_\SO = - \frac{e^2}{\pi\hbar} \frac{L}{(2\pi)^2}
 \sum_{\alpha,\beta = \pm 1}
 \braopket{x,\,\alpha,\,\beta}
          {\frac{1}{\gamma - h^C_{\SO}} }
          {x,\,\beta,\,\alpha}   .
\end{equation}
With Eq.~(\ref{eqnHSOC}), we obtain the magnetoconductance

\ifpreprintsty\else\end{multicols}\emcsep\fi

\begin{eqnarray}
\label{eqnMcSOExact}
\dgMC_\SO &=& 
-\frac{e^2}{\pi\hbar} \sum_{j=-\infty}^\infty
{ \ifpreprintsty\textstyle\else\fi
\frac{
2  \left[4\kappa^2 + \left( m^2 + \gamma  \right) ^2 \right]  
   \left(m^2 + \gamma +1 \right)  
+S^2 \left[8m^4 + 2m^2 \left(4 \gamma -1 \right) + 2 \gamma + 1  \right]
}
{
  \left[4\kappa^2 + \left( m^2 + \gamma  \right) ^2 \right] 
   \left[m^4 + 2 m^2 
     \left(\gamma -1 \right)  + \left( \gamma +1  \right)^2 \right]  + 
      S^2 \left( m^2 + \gamma  \right)  \left[ 
           4 m^4 + m^2 \left(4 \gamma -3\right) + \gamma + 1   \right] 
}
},
\end{eqnarray}

\ifpreprintsty\else\bmcsep\begin{multicols}{2}\fi

\noindent
where $m=j-2\phi/\phi_0$ contains the Aharonov-Bohm flux.
In Sec.~\ref{secPS} we will see that in 
 the ``adiabatic'' limit $\kappa,\,S \gg 1$
 the magnetoconductance $\dgMC_\SO$ will show some
 similar properties as for inhomogeneous fields,
 in particular a peak-splitting in the power spectrum,
 see Fig.~\ref{figMcSoPS}.

\subsection{Conductance Fluctuations}
\label{ssecSOUCF}

We turn now to a discussion of 
 the recent experiment by Morpurgo et~al.\cite{morpurgo}
 by specifying the parameters of 
 the effective Hamiltonian $h^{C/D}_\SO$,
 as given in
 Eqs.~(\ref{eqnHSO}), (\ref{eqnHSOC}), and (\ref{eqnHSOD}).
In Ref.~\onlinecite{morpurgo}, conductance measurements were performed
 on an InAs ring, with nearly ballistic transport.
For the parameters given, \cite{morpurgo}
 $\alpha = 5.5 \times 10^{-10}\:{\rm eV\, cm}$,
 $L      = 6.6                \:\mu{\rm m}$,
 $\vfermi= 9.8 \times 10^{7}  \:{\rm cm}/{\rm s}$,
 $\ell   = 1.0                \:\mu{\rm m}$, and
 $D      = \vfermi \ell/2 = 4.9 \times 10^{3} \:{\rm cm}^2/{\rm s}$,
we calculate with Eq.~(\ref{eqnSOCoupling}) 
 a numerical value of $S\approx 1/50$.
Compared to this, the strength of the Zeeman term $\kappa\approx1/2$
 (with $|g|=15$) is much larger.
Within the diffusive approximation,
 this spin-orbit coupling $S\ll\kappa$
 gives only a negligible contribution to the 
 effective Hamiltonian $h^{C/D}$ [Eq.~(\ref{eqnHSO})]
 and thus does not produce any Berry phase effects.
This very same finding has also been obtained in Ref.~\onlinecite{malshukovSO},
 based on a slightly different reasoning.
Still, we show in Sec.~\ref{secPS} that a spin-splitting 
 produced by spin-orbit interaction can be obtained
 in the ``adiabatic regime'' $\kappa,\,S\gg 1$,
 which, however, is in the opposite limit to the one 
 reported in Ref.~\onlinecite{morpurgo}.
So although we cannot give a quantitative explanation of 
 the experiment\cite{morpurgo} here, 
 we can offer a qualitative interpretation, see Fig.~\ref{figSoUcf2D}.
Further, there is an uncertainty in the spin-orbit coupling parameter $\alpha$
 in InAs, as it was recently pointed out\cite{brosig_ensslin},
 and more experiments might be needed to clarify this issue.

\begin{figure}[H]
\centerline{\psfig{file=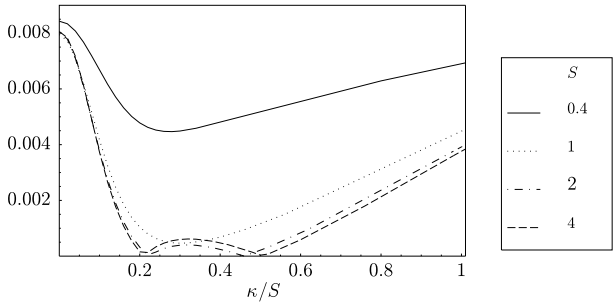,width=\figwidth}}
\caption{ The normalized
 amplitudes of the $h/2e$ oscillations of the UCFs
 with spin-orbit coupling, $\dgUCF_\SO$.
The power spectrum of the Aharonov-Bohm oscillations
 was calculated at different
 values $\kappa=\tilde{\kappa}$ of the perpendicular fields by
 varying the Aharonov-Bohm flux $0\leq \phi=\tilde{\phi}\leq 1$.
 From the power spectrum, 
 the frequency contribution of the $h/2e$ oscillation
 was normalized by the zero frequency contribution
 and is shown here as a function of $\kappa/S$.
We have assumed $T=0$ and $\gamma^C=\gamma^D=0.1$.
}
\label{figUcfSoAmpl}
\end{figure}

 To this end we calculate the exact,
 i.e.\ without assuming any form
 of adiabaticity,
 expression for the conductance fluctuations $\dgMC_\SO^{(2)}$
 in the presence of spin-orbit interaction.
With the block-diagonalization of the
 Hamiltonian $h^{C/D}_\SO$ [Eqs.~(\ref{eqnHSOC}), (\ref{eqnHSOD})]
 we obtain the propagators
 required in the formula for the conductance correlator 
 [Eq.~(\ref{eqnUcfTrChi})].
We use Mathematica to obtain an explicit algebraic expression
 for $\dgUCF_\SO$ (which is lengthy and thus not reproduced here)
 and plot it in Fig.~\ref{figUcfSoAmpl}
 (see also Figs.~\ref{figSoUcfH2eOsc} and~\ref{figSoUcf2D}).
 From this plot
 we deduce that in a configuration with spin-orbit coupling,
 the Aharonov-Bohm oscillations vanish for certain values of $S$ and $\kappa$.
It is remarkable that this happens, for $S \geq 2$, at
 the fixed ratios $\kappa/S=0.2$ and $0.5$,
 which can be ascribed again to some effective magic angles.
Thus we see that Berry phase-like effects occur in $\dgUCF_\SO$
 as the amplitudes of the Aharonov-Bohm oscillations become dependent
 on $\kappa/S$.
This resembles the case for inhomogeneous fields, 
 where the amplitudes of the Aharonov-Bohm oscillations
 became dependent on the tilt angle $\eta$ of the magnetic field
 due to the Berry phase,
 as it was shown in Sec.~\ref{ssecQual}.

\section{Peak Splittings in Power Spectra}
\label{secPS}

\subsection{Frequency Shifts in $\dgMC$ and $\dgUCF$}
\label{secSplitLimits}

We discuss now the emergence of the Berry phase in terms of
 a splitting of the frequencies 
 of the Aharonov-Bohm oscillations
 in the magnetoconductance\cite{LSG,Stern} $\dgMC$
 and in the UCFs\cite{LSG} $\dgUCF$,
 which can be made visible in the power spectrum.\cite{morpurgo}
Both quantities depend on the spin-dependent total phase
 $\Phi_{\alpha}$, given
 here for the special case of the texture defined in Eq.~(\ref{eqnField})
 and for two equal fields $\vec{B}=\vec{\tilde{B}}$,
\begin{eqnarray}
\label{eqnPhaseApprox}
\Phi_{\pm1} & = &
 2\phi/\phi_0 \pm \cos{\eta}
 = 2\phi/\phi_0 \pm \frac{1}{\sqrt{1+(B_\parallel/B_z)^2}}
\nonumber \\ & %noPP
 \approx 
 &  %noPP
2 \phi/\phi_0 \pm B_z/B_\parallel
 =  B_z \left(2 B_{\phi_0}^{-1} \pm B_\parallel^{-1} \right)
.
\end{eqnarray}
The approximation used here is valid for small 
 perpendicular fields $B_z \ll B_\parallel$.
We have introduced $B_{\phi_0} = \phi_0/A$ as the perpendicular
 field which produces a flux of one flux quantum $\phi_0$ through the ring,
 i.e.\ the period of an Aharonov-Bohm oscillation in $\phi$.
The Berry phase is not sensitive to the area enclosed by the ring;
 thus we prefer here to describe oscillations in $B_z$ rather than in $\phi$.
As both $\dgMC$ and $\dgUCF$
 contain periodic terms in $\Phi_{1}$ and $\Phi_{-1}$,
 they exhibit oscillations in $B_z$
 with the Aharonov-Bohm frequency for homogeneous fields, $2B_{\phi_0}^{-1}$,
 shifted (at $B_z=0$) by the frequency
\begin{equation}
\label{eqnFreqShiftI}
 \frac{1}{\Delta B^{}_{1}} = \pm \frac{1}{B_\parallel}
 \,,
\end{equation}
  which results in a peak splitting in the power spectrum.

These splittings are, however, generally on the order of the
 resolution of the spectrum, which makes it difficult to make them visible.
If the perpendicular field is varied from $-B_{\rm max}$ to $B_{\rm max}$,
 the discrete Fourier transform (DFT) of such an interval has a
 resolution of $1/2B_{\rm max}$,
 i.e.\ the sampling frequencies are separated by this value.
Thus, the peak-splitting term can only be made visible if this resolution 
 is high enough, i.e.\ $1/2B_{\rm max} \leq 1/B_\parallel$, or

\begin{equation}
 B_{\rm max} \geq \frac{1}{2} B_\parallel
 .
\end{equation}

We note that this restriction is still consistent with
 the approximation made in Eq.~(\ref{eqnPhaseApprox}),
 since for $B_z=B_\parallel/2$ the
 approximated value of the Berry phase is larger than the exact value
 by only a factor of $\sqrt{5}/2 \approx 1.1$.

Now we consider the case beyond the above approximation.
Here, an estimate for the frequency shifts can be obtained
 by counting the additional oscillations
 upon increasing $B_z$.
In this estimation we again neglect the change in frequency 
 of the Aharonov-Bohm oscillations while $B_z$ is increased.
However, now we take the mean value of the frequency instead
 of the frequency at $B_z=0$ as in Eq.~(\ref{eqnFreqShiftI}).
Varying $B_z$ from $0$ to $B_{\rm max}$ changes the Berry phase contribution
 to $\Phi_{\pm1}$ [Eq.~(\ref{eqnPhaseApprox})]
 from $0$ to $\pm\cos{\eta|_{B_z=B_{\rm max}}}$,
 and so we obtain the mean frequency shift

\begin{eqnarray}
\label{eqnFreqShiftII}
 \frac{1}{\Delta B^{}_{2} } &=& 
 \pm \frac{1}{\sqrt{B_{\rm max}^2 + B_\parallel^2}}
%\nonumber \\
  \approx 
 \pm \frac{1}{B_\parallel} \left(1 - \frac{B_{\rm max}^2}{2 B_\parallel^2}
 \right).
\end{eqnarray}
When we have calculated the DFT of $\dgMC$ and $\dgUCF$,
 we have confirmed the predictions given above,
 i.e.\ we do not observe a peak splitting in the $2 B_{\phi_0}^{-1}$ frequency
 for low $B_{\rm max}$, due to an insufficient resolution of the DFT.
However, we do see a peak splitting in the DFT for higher fields 
 (see Figs.~\ref{figMcOsc},~\ref{figUcfH2eOsc}),
 which vanishes again for $B_{\rm max} \gg B_\parallel$.
Since studies of the DFT suffer from
 a restricted resolution,
 it might be more promising to search for the Berry phase via the effects
 discussed in Sec.~\ref{ssecQual}.

Finally, we point out that an anisotropic $g$ 
 factor affects the size of the frequency splitting.
If the $g$ factor perpendicular to the ring, $g_z$, is larger
 than the one in the plane of the ring, $g_\parallel$,
 the Berry phase dependence on $B_z$ increases
 while the Aharonov-Bohm phase remains unaffected.
As the total phase is
 $\Phi_{\pm1} \approx 2\phi/\phi_0 \pm g_z B_z/g_\parallel B_\parallel$,
 the frequency splitting is increased
 by a factor of  $g_z/g_\parallel$.

\subsection{Frequency shifts in $\dgUCF_\homtag$ for homogeneous fields}
\label{ssecFreqShiftHomFields}

At this point it is important to realize that frequency shifts can also
 appear in the conductance fluctuations $\dgUCF$ for homogeneous fields,
 i.e.\ even when there is no Berry phase present.
For homogeneous fields the evaluation of Eq.~(\ref{eqnUcfTrChi})
 is straightforward, as $h^{C/D}$ [Eq.~(\ref{eqnHMatrix})] becomes diagonal,
 see also App.~\ref{secUcfLimits}.
We evaluate the DOS terms, i.e.\ the terms containing
 ${\rm Re}\,{\rm Tr}\, \hat{\chi}_\omega \hat{\chi}_\omega$ 
 in Eq.~(\ref{eqnUcfTrChi}),
 in the low temperature limit
 for $\eta=\tilde{\eta}=0$:

\begin{eqnarray}
\label{eqnUcfHom}
 & & \dgUCF_{DOS, \atop C/D}   \propto
  {\rm Re} \!\! \sum_{j \atop \alpha,\tilde{\alpha}=\pm1}
  \frac{1}{\left[\gamma + (j \!-\!\Phi^{\scriptscriptstyle C/D})^2
           + i (\alpha \kappa\!+\!\tilde{\alpha} \tilde{\kappa}) \right]^2 }
\\
 & & \!\! \approx \frac{2 \pi}{\gamma^{3/2}} +
  \!\sum_{\alpha, \tilde{\alpha}} \sum_{n=1}^{\infty}  \! \frac{2\pi^2 n}{\gamma}
 e^{-2\pi n \sqrt{\gamma} } \cos{\!\left[\!2\pi n \!
   \left(\!\Phi^{\scriptscriptstyle C/D} \!\!+\! 
   \frac{\alpha\kappa\!+\!\tilde{\alpha}\tilde{\kappa}}{2\sqrt{\gamma}} \right)\!\right]}
  ,
\nonumber 
\end{eqnarray}
where we have defined 
 $\Phi^{C/D} = \phi/\phi_0 \pm \tilde{\phi}/\phi_0$.
The approximation on the second line of Eq.~(\ref{eqnUcfHom}) is valid
 for $\gamma \gg 1/4\pi^2,\: \alpha\kappa+\tilde{\alpha}\tilde{\kappa}$.
 From Eq.~(\ref{eqnUcfHom}),
we see that the Zeeman term itself already leads to  a frequency splitting.
So, for instance,
 if we take the Fourier transform of $\dgUCF(B_z,\,-B_z)$ with respect to
 $B_z$, we can observe a frequency splitting of the $h/e$ oscillations of
 the diffuson contribution in the DOS term $\dgUCF_{DOS,\,D}$, given by
\begin{equation}
\label{eqnFreqShiftHom}
\frac{1}{\Delta B_{\mathit Zeeman}} =
  \pm \frac{g \mubohr}{4 \hbar D} \frac{L_D L}{2\pi}
 .
\end{equation}
We checked numerically that the estimated frequency splitting
 [Eq.~(\ref{eqnFreqShiftHom})]
 is correct within 20 percent
 even for parameters beyond the assumptions made for the second
 line of Eq.~(\ref{eqnUcfHom}).
It is important to keep this property of the conductance fluctuations $\dgUCF$
 in mind, when searching for Berry phase effects.
If vanishing Aharonov-Bohm oscillations 
 or peak splittings in the power spectrum are used to identify 
 the presence of a Berry phase,
 one has to rule out effects coming from the Zeeman term in the UCFs,
 e.g.\ by comparison with the results for homogeneous fields.

\begin{figure}
\centerline{\psfig{file=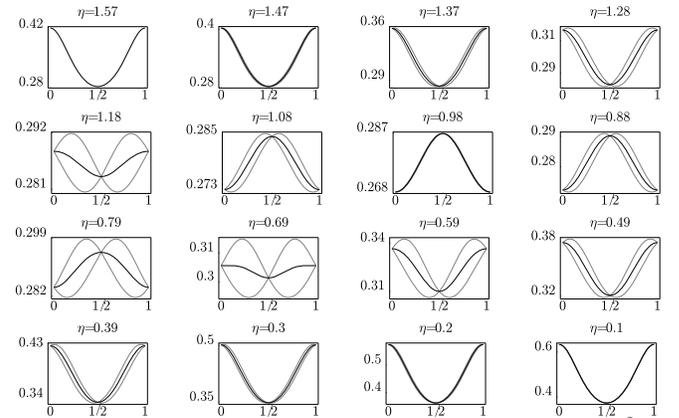,width=\figwidth}}
\caption{ The magnetoconductance $\dgMC$ in units of $-2e^2/h$
 as a function of the Aharonov-Bohm flux $2\phi/\phi_0$,
 for different tilt angles $\eta$ of the external field.
 We have chosen the dephasing $\gamma=0.1$ and
 the field $B_\parallel$ parallel to the ring plane 
 to be constant, defined through
 $B_\parallel \propto \kappa_\parallel = \kappa\,\sin{\eta}=2.0$.
The magnetoconductance is shown in black, while its contribution
 from the different spins $\alpha=\pm1$ 
 are scaled by a factor of two and drawn in gray.}
\label{figMcAB}
\end{figure}

\subsection{Numerical Evaluations}
\label{ssecPSTechniques}

We shall now numerically evaluate the  magnetoconductance $\dgMC$
 for a ring in an inhomogeneous field.
We base our analysis on the calculations from
 Ref.~\onlinecite{LSG99}.
In Fig.~\ref{figMcAB} we show the Aharonov-Bohm oscillations
 for different tilt angles $\eta$ of the external field $\vec{B}$, 
 which is
 set so strong that we are well within the adiabatic regime.
We can readily see that for $\eta\approx \pi/3$ 
 a phase shift of $\pi$ occurs,
 which comes directly from the Berry phase,
 compared to the oscillations at $\eta=0$ and $\eta=\pi/2$.
For the intermediate tilt angles the effect of the Berry phase is 
 only visible
in the amplitude of the Aharonov-Bohm oscillations,
 as the phase shifts for the two spin directions occur
 with opposite signs and thus---if
 both spin directions contribute equally---no phase-shift effect is visible.

As such a phase shift at $\pi/3$ might not be easy to observe,
 studying signs in the power spectrum provides an 
 interesting alternative,\cite{morpurgo}
 even though it requires a sufficiently high resolution, 
 as discussed in Sec.~\ref{secSplitLimits}.
Indeed, we can observe a peak splitting in the spectrum of the
 magnetoconductance, as shown in the inset of Fig.~\ref{figMcOsc}.
We notice an even more distinct feature: the Aharonov-Bohm oscillations
 vanish at two magic tilt angles, 
 $\cos{\eta} = 0.4, \, 0.75$, of the field.
The mechanism for this effect is exhibited in
 Fig.~\ref{figMcAB},
 where it is shown how the two contributions of the different spins
 suppress the oscillations.

\begin{figure}[H]
\centerline{\psfig{file=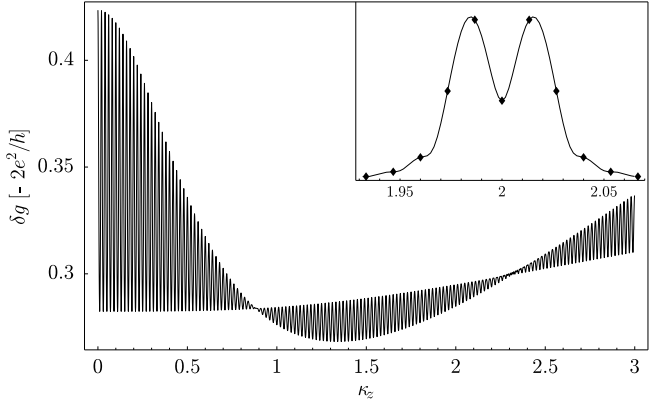,width=\figwidth}}
\caption{ The Aharonov-Bohm oscillations in the magnetoconductance $\dgMC$
 as a function of the perpendicular field $B_z$,
  shown here as $\kappa_z = \kappa\,\cos{\eta}$.
The radial field component has a magnitude of 
 $B_\parallel \propto \kappa_{\parallel} = \kappa\,\sin{\eta} = 2.0$
 and $\gamma=0.1$.
The vanishing oscillations near 
 $\kappa_z \approx 0.9 ,\, 2.3$
(for the magic angle $\cos{\eta} \approx 0.4, \, 0.75$)
 are striking;
 this a direct consequence of the Berry phase,
 arising from a canceling of the oscillating
 contributions of opposite spin directions.
The inset shows the power spectrum\protect\cite{gtte} where a peak splitting is
 visible. }
\label{figMcOsc}
\end{figure}

At this point, we would like to stress that the peak
 splitting depends strongly on the different dephasing terms.
In particular, one cannot rely on calculations where the dephasing due to
 the inhomogeneous fields is not properly taken into account.
So if the dephasing $\gamma$ due to homogeneous effects is very small,
 e.g.\ on the order of $1/100$,
 the amplitude of the oscillations gets reduced drastically as soon
 as the tilt angle $\eta$ changes from $\pi/2$
 to a smaller, nonzero value,
 since the field inhomogenity causes additional dephasing.
Thus the Fourier transform of such oscillations 
 has a dominant contribution only from
 the first few oscillations close to $\pi/2$.
This suppression of the remaining oscillations 
 acts as a narrowing of the data window\cite{nr}
 and leads to a widening of the peaks in the
 power spectrum, masking the peak splitting.
The oscillations are further suppressed by the additional 
 dephasing arising from an increasing perpendicular field, which 
 penetrates the ring arms.
Of course, it is possible to remove this unwanted over-emphasizing of
 certain oscillations from experimental data
 in a post-processing step;
 using a standard windowing function
 (we used the Hann window\cite{nr} for the inset of Fig.~\ref{figMcOsc})
 for DFTs
 greatly reduces this problem,
 in addition to the usual reduction
 of components leakage of neighboring frequencies
 in the power spectrum.\cite{nr}

\begin{figure}[H]
\centerline{\psfig{file=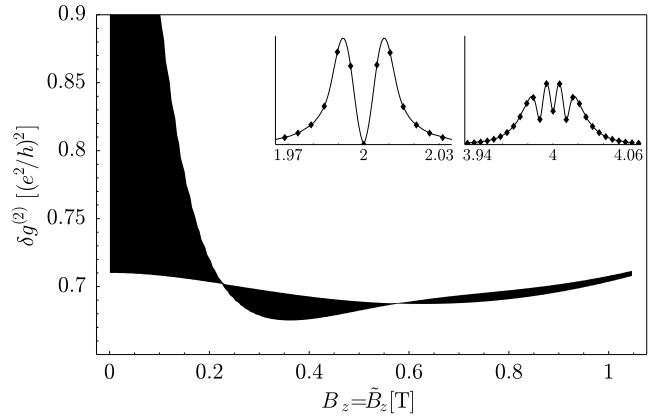,width=\figwidth}}
\caption{ The UCFs $\dgUCF$ for $\vec{B}=\vec{\tilde{B}}$
 plotted as function of $B_z$ (see first part of Sec.~\ref{ssecQual}).
 While the printing resolution is not high enough to show
 the Aharonov-Bohm oscillations,
 the envelope clearly illustrates the non-monotonic behavior of
 their amplitudes,
 which vanish at the magic angles $\eta=0.72, 1.15$.
 We have taken  a fixed radial component for both fields of
 $B_\parallel = \tilde{B}_\parallel = 0.5 \:{\rm T}$.
We have assumed $L = 3\: \mu{\rm m}$,
 $D=65\:{\rm cm}^2/{\rm s}$, and $T=0$.
The dephasing was taken into account according to Eq.~(\ref{eqnLCD}),
 with the parameters 
 $L_\varphi = 1.5\: \mu{\rm m}$,
 and $a= 60 \:{\rm nm}$.
The two insets show the contributions of the $h/2e$ and $h/4e$
 oscillations
 to the power spectrum\protect\cite{gtte} in arbitrary units plotted
 against the frequency in units of $\phi_0^{-1}$.
The right inset was scaled
 by a factor of 10.
For the particular range of $B_z$ chosen here,
 there is a peak splitting visible for the $h/2e$ oscillations,
 while 
 we observe four peaks around the $h/4e$ frequency.
 }
\label{figUcfH2eOsc}
\end{figure}

For the conductance fluctuations $\dgUCF$,
we will further illustrate the effects of the two configurations discussed
 in Section~\ref{ssecQual}.
In Fig.~\ref{figUcfH2eOsc} we show the Aharonov-Bohm oscillations occurring
in $\dgUCF$
 when the fields are equal, i.e.\ $\vec{B} = \vec{\tilde{B}}$.
Taking the discrete Fourier transform of $\dgUCF$
 over the range $B_z=0,\dots,1\:{\rm T}$, yields a clear
 peak splitting of the contribution of the $h/2e$ oscillations to the power
 spectrum, see left inset in Fig.~\ref{figUcfH2eOsc}.
We notice a splitting into four peaks of the contribution
 of the $h/4e$ oscillations 
 (right inset of Fig.~\ref{figUcfH2eOsc}).
They only occur in the exact solution $\dgUCF$,
 whereas $\dgUCF_\adtag$ exhibits only two peaks
 if we ignore the $\eta$, $\tilde{\eta}$-dependent dephasing,
 i.e.\ set $\gammatilde\to\gamma^{C/D}$ and $P\to 0$ in Eq.~(\ref{eqnUCFad}).
We point out that
 the frequency shifts for the $n$th harmonics of the Aharonov-Bohm oscillations
 increase with~$n$ 
 and are thus are better resolved in the power spectrum
 with increasing~$n$.

We plot $\dgUCF(\vec{\tilde{B}})$ in Fig.~\ref{figUcfHeOsc}
 for the special case $\vec{\tilde{B}}=(0,\,0,\,\tilde{B}_z)$ homogeneous
 (see also Sec.~\ref{ssecQual}).

Finally, we consider the power spectrum of the magnetoconductance
 $\dgMC_\SO$
 in the presence of spin-orbit coupling.
We use Eq.~(\ref{eqnMcSOExact}) and ignore for simplicity dephasing due
 to the external magnetic fields penetrating the arms of the ring.
Indeed, taking the Fourier transform of the magnetoconductance,
 a spin splitting can be observed.
However, the splitting is not as pronounced as in the case for inhomogeneous
 fields.
Especially important, the splitting is only visible for sufficiently large dephasing
 parameters $\gamma$ (produced by inelastic scattering),
 which can be seen in Fig.~\ref{figMcSoPS}.
In contrast to the effects discussed before, using a windowing function
 was not sufficient to identify a peak splitting for moderately small
 dephasing parameters $\gamma \lesssim 0.3$. 
Qualitatively, however, the power spectra of the magnetoconductance 
 for inhomogeneous magnetic fields and for spin-orbit coupling agree,
 with both showing a peak splitting.

\begin{figure}[H]
\centerline{\psfig{file=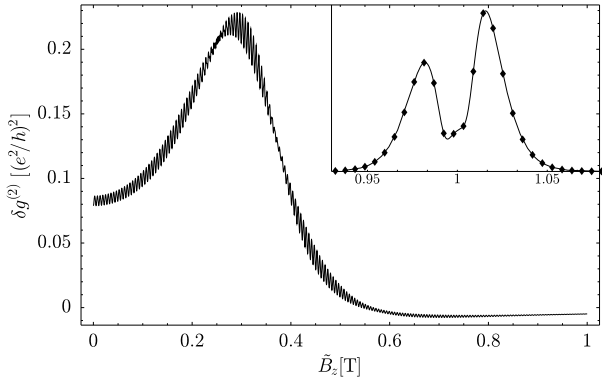,width=\figwidth}}
\caption{ The UCFs $\dgUCF$ for
 a homogeneous texture of $\vec{B}$
 plotted as function of $\tilde{B}_z$  
 (see second part of Sec.~\ref{ssecQual}).
 We have taken the homogeneous field as
 $B_z = 0.5 \:{\rm T}$, and $B_\parallel = 0\:{\rm G}$
 and have fixed the radial component for the other field as
 $\tilde{B}_\parallel = 0.43 \:{\rm T}$.
The remaining parameters are chosen as in Fig.~\ref{figUcfH2eOsc}.
The inset shows the power spectrum\protect\cite{gtte}
 in arbitrary units plotted
 against the frequency in units of $\phi_0^{-1}$,
 which exhibits a splitting in the $h/e$ contributions.
 }
\label{figUcfHeOsc}
\end{figure}

\begin{figure}[H]
\centerline{\psfig{file=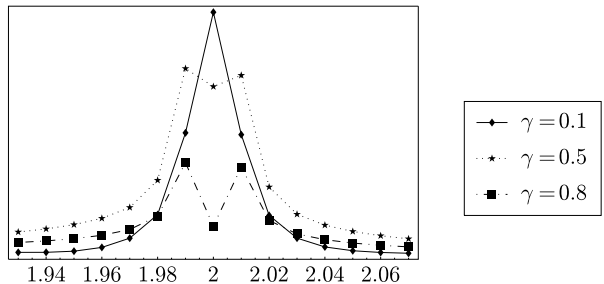,width=\figwidth}}
\caption{ The power spectrum of the magnetoconductance $\dgMC_\SO(B)$
 with spin-orbit coupling, Eq.~(\ref{eqnMcSOExact}),
 in arbitrary units plotted
 against the frequency in units of $\phi_0^{-1}$.
 We have chosen $S=4$ 
 and taken the Fourier transform of the magnetoconductance
 for $0\leq\kappa\leq 4$.
 We show the power spectrum for three different values
 of the dephasing parameter $\gamma$,
 where we have downscaled the values for $\gamma=0.1$ by a factor of $10$.
Note that a peak splitting occurs only for the cases with larger dephasing.
}
\label{figMcSoPS}
\end{figure}

The UCFs with spin-obit interaction $\dgUCF_\SO$ are plotted in
 Fig.~\ref{figSoUcfH2eOsc} as a function of the perpendicular fields
 $B_z = \tilde{B}_z$.
 We observe a Berry phase-like frequency splitting in the power spectrum.
However, as this splitting is rather small, it is only visible in
 the $h/4e$ oscillations, 
 where the splitting is twice as large as in the $h/2e$ oscillations.
Again, the suppression of the Aharonov-Bohm oscillations at
 $\kappa/S \approx 0.25$ is a distinct feature of a Berry phase-like effect.

\begin{figure}[H]
\centerline{\psfig{file=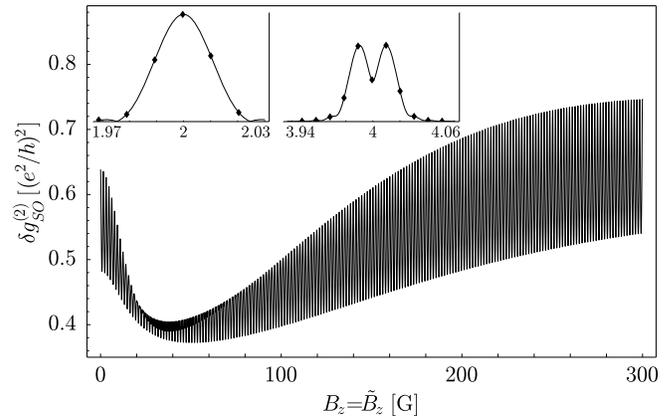,width=\figwidth}}
\caption{ The UCFs $\dgUCF_\SO$ with spin-orbit interaction
 for $\vec{B}=\vec{\tilde{B}}$ plotted as function of $B_z$.
We have taken
 $\alpha = 1.0 \times 10^{-9}\:{\rm eV\, cm}$,
 $L      = 12.5                \:\mu{\rm m}$,
 $D      = 2.0   \times 10^{-2} \:{\rm m}^2/{\rm s}$,
 $g      = 15$,
and have assumed $T=0$.
This gives us $S=1.6$ [Eq.~(\ref{eqnSOCoupling})], and
 $\kappa(B_z = 300\:{\rm G}) = 4.2$ [Eq.~(\ref{eqnKappa})]. 
The dephasing was taken into account according to Eq.~(\ref{eqnLCD}),
 with the parameters 
 $L_\varphi = 5.0\: \mu{\rm m}$,
 and $a= 120 \:{\rm nm}$.
The envelope of the Aharonov-Bohm oscillations shows a non-monotonic
 behavior, which also appears in the UCFs for inhomogeneous fields $\dgUCF$
 (see Fig.~\ref{figUcfH2eOsc}).
The $h/2e$ oscillations are strongly suppressed
 at $B_z \approx 30 \:{\rm G}$, which corresponds to $\kappa/S \approx 0.25$,
 as can also be seen from Fig.~\ref{figUcfSoAmpl}.
However, this suppression is not very obvious in this figure,
 since $h/4e$ oscillations are present for $B_z \approx 30\:{\rm G}$.
The two insets show the contributions of the $h/2e$ and $h/4e$
 oscillations to the power spectrum in arbitrary units\protect\cite{gtte}
 plotted
 against the frequency in units of $\phi_0^{-1}$.
The right inset was scaled
 by a factor of 10.
For the particular range of $B_z$ chosen here,
 there is only a single peak visible for the $h/2e$ oscillations,
 while we observe a small peak splitting around the $h/4e$ frequency.
 }
\label{figSoUcfH2eOsc}
\end{figure}

A quantity, which was subject of recent studies,\cite{morpurgo,malshukovSO}
 is the disorder-averaged squared power spectrum of the conductance
\begin{equation}
\label{eqnGnuAbs}
 \Big\langle \big|g(\nu)\big|^2 \Big\rangle 
   = 
 \big| \big\langle g(\nu) \big\rangle \big|^2 +
 \Big\langle \big| g(\nu) -\langle g(\nu)\rangle \big|^2 \Big\rangle
.
\end{equation}
On the one hand, we recognize that
 the first term contains the Fourier transform of the
 (averaged) magnetoconductance $\dgMC$, 
 which has frequency contributions from its $h/2e$ oscillations.
On the other hand, the second term of Eq.~(\ref{eqnGnuAbs}) 
 is given through the conductance fluctuations $\dgUCF$ as
 $\int\int d B_z d\tilde{B}_z \exp{\{2\pi i \nu (B_z-\tilde{B}_z)\}}
    \dgUCF(\vec{B},\vec{\tilde{B}})$.
This term contributes
 frequencies corresponding to $h/e$ oscillations, coming from
 the diffuson term $\dgUCF_D$ in the conductance fluctuations.
Thus, if we now investigate $h/e$ oscillations, we can restrict our
 studies to the second term of Eq.~(\ref{eqnGnuAbs}).
We have evaluated $\langle|g(\nu)|^2\rangle$ for inhomogeneous fields, with
 the parameters given in the caption of Fig.~\ref{figUcfH2eOsc}.
A splitting of the frequency corresponding to the $h/e$ oscillations was
 observed and was identified not to result from the Berry phase but
 from the Zeeman term already present in the case of homogeneous fields
 [Eq.~(\ref{eqnFreqShiftHom})].
Then we examined $\langle|g_\SO(\nu)|^2\rangle$ with spin-orbit coupling
 for various parameters.
An additional peak splitting to the one produced by the 
 Zeeman term [Eq.~(\ref{eqnFreqShiftHom})] appears 
 for some specific parameters,
 i.e.\ for $S$ large enough to reach ``adiabaticity'' and
 for large enough sampling intervals of $B_z$, $\tilde{B}_z$ to
 obtain a sufficiently high resolution in the power spectrum.
In Fig.~\ref{figSoUcf2D} we see such a splitting of the $h/e$ contribution
 into four peaks.
However, using the parameters given in Ref.~\onlinecite{morpurgo},
 we have $S\approx 1/50$ and $\kappa\approx1/2$ (see Sec.~\ref{ssecSOUCF})
 and in this regime we do not observe any peak splitting, in accordance
 with Ref.~\onlinecite{malshukovSO}.

\begin{figure}[H]
\centerline{\psfig{file=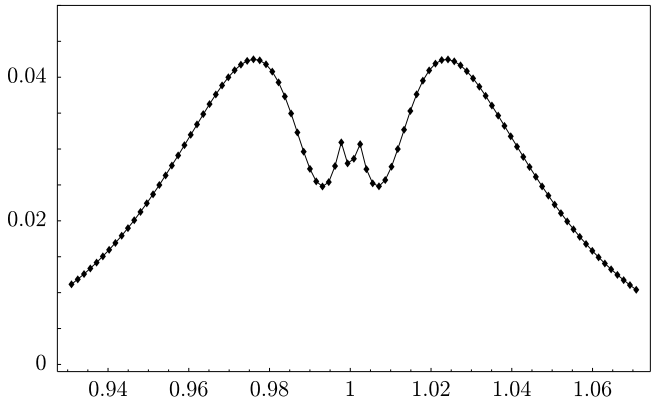,width=\figwidth}}
\caption{ The disorder-averaged squared power spectrum of the conductance
 $\langle| g_\SO(\nu) |^2\rangle$ [Eq.~(\ref{eqnGnuAbs})]
 with spin-orbit interaction
 plotted as function of $\nu$ in units of $\phi_0^{-1}$, 
 normalized by the zero frequency contribution.
We have taken the same parameters as in Fig.~\ref{figSoUcfH2eOsc}, but
 now with
 $\alpha = 2.0 \times 10^{-9}\:{\rm eV\, cm}$,
 and thus $S=3.2$ [Eq.~(\ref{eqnSOCoupling})]. 
We have calculated the second term of Eq.~(\ref{eqnGnuAbs}) explicitly
 (see text),
 while taking $B_z,\, \tilde{B}_z \in [-B_{\rm max}, \, B_{\rm max}]$ with
 $B_{\rm max} = 0.1 \: {\rm T}$,
 which gives us a maximal value
 $\kappa_{\rm max} = 14$ [Eq.~(\ref{eqnKappa})].
The peak splitting into the inner two peaks is produced by the spin-orbit
 interaction, while the larger satellite peaks result from the Zeeman term
 [Eqs.~(\ref{eqnUcfHom}) and (\ref{eqnFreqShiftHom})].
}
\label{figSoUcf2D}
\end{figure}

\section{Conclusion}
We have calculated the exact conductance fluctuations $\dgUCF$
 for a special texture [Eq.~(\ref{eqnField})]
 and given its adiabatic approximation $\dgUCF_\adtag$.
In addition to the already known differential equations for the cooperon
 we have derived the ones for the diffuson in inhomogeneous magnetic fields
 (App.~\ref{secCdDiffEqn}).
With the result $\dgUCF_\adtag$ the dephasing
 due to inhomogeneous fields became explicit and could be compared with
 previous calculations\cite{LSG} where adiabatic eigenstates were used
 and this dephasing was only implemented with a phenomenological parameter.
Then we have described some magic tilt angles of the magnetic field
 at which the Berry phase suppresses the Aharonov-Bohm oscillations.
We have used this effect to illustrate how 
 the adiabatic criterion becomes gradually satisfied.
We have calculated numerically the required magnetic field strength
 for which the adiabatic approximation becomes valid 
 and have 
 shown that the adiabatic criterion
 is less stringent for diffusive than for ballistic motion,
 thus confirming previous findings.\cite{LSG,LSG99}

Furthermore, we have calculated the magnetoconductance and
 the conductance fluctuations  for
 a diffusive conductor in the presence of spin-orbit coupling.
A numerical analysis revealed
 a non-monotonic behavior of the amplitudes of the Aharonov-Bohm oscillations
 and peak-splittings in the power spectrum---observations that are
 similar to the Berry phase effects we have found for
 inhomogeneous magnetic fields.

Finally, we have described the mechanisms 
 which lead to peak splittings
 in the power spectrum of magnetoconductance and UCFs
 and have discussed numerical requirements to make such 
 peaks splittings visible.

\acknowledgments
\addcontentsline{toc}{section}{Acknowledgments}
We would like to thank
 G. Burkard,
 R. H\"aussler,
 and E.V. Sukhorukov
 for many fruitful discussions.
This work has been supported in part by the Swiss National Science Foundation.

\ifpreprintsty\else\end{multicols}\emcsep\fi

\appendix

\section{Differential equations for Cooperon and Diffuson}
\label{secCdDiffEqn}
Here we will transform the exact conductance correlator
 for diffusive systems and arbitrary magnetic textures 
 to make a Schr\"odinger equation approach \cite{LGlong} possible.
Further we will derive the explicit differential equation 
for the diffuson propagator (the one for the cooperon  has been derived
previously~\cite{LSG}).

The conductance correlator has been derived in Ref.~\onlinecite{LSG},
using diagrammatic techniques, and is given by
\begin{eqnarray}
\label{ucfChi}
\dgUCF &  =   &
\left(\frac{2 e^2 D}{h L^2}\right)^2  
  \int d\epsilon \, d\epsilon' n'(\epsilon) \, n'(\epsilon') 
  \int d\vec{x} \, d\vec{x}'
  \sum_{\alpha_1,\alpha_2,\alpha_3,\alpha_4}
  \left\{ 
   \frac{1}{d} \left|
    \chi^C_{\alpha_1\alpha_2,\alpha_3\alpha_4}(\vec{x},\vec{x'},\omega)
   \right|^2 \right.
\nonumber \\
 & + & \left.
 2\, {\rm Re} \left[
    \chi^C_{\alpha_1\alpha_2,\alpha_3\alpha_4}(\vec{x},\vec{x'},\omega) 
    \chi^C_{\alpha_2\alpha_1,\alpha_4\alpha_3}(\vec{x'},\vec{x},\omega) \right]
  + \left[ \chi^C \rightarrow \chi^D \right]
   \right\}
 ,
\end{eqnarray}
where $n'(\epsilon)$ is the derivative of the Fermi function,
$\hbar\omega=\epsilon-\epsilon'$,
and $d$ describes the dimension of the system with respect to 
the mean free path $l$.
The
inverse Fourier transform of the cooperon/diffuson propagators
$\chi^{C/D}(\vec{x'},\vec{x},w)$ were obtained~\cite{LSG,Schmid} as
\begin{eqnarray}
 \chi^{C/D}_{\alpha_1 \alpha_2,\alpha_3 \alpha_4}(\vec{x'},\vec{x};t',t)
  & = & \theta(t'-t) \int_{\vec{R}(t)=\vec{x}}^{\vec{R}(t')=\vec{x'}} 
           {\cal D}\vec{R}
       \exp\left\{ -\frac{1}{4 D} \int_t^{t'} d \tau |\vec{\dot{R}}|^2 \right\}
       \nonumber\\
  & \times & \exp \left\{i \frac{e}{\hbar} \int_t^{t'}d \tau \: \left[
       \vec{\dot{R}} \cdot \vec{A}^{\rm em}(\vec{R}(\tau)) + 
       \vec{\dot{R}}^{\pm}\cdot \vec{\tilde{A}}^{\rm em}(\vec{R}^{\pm}(\tau))
      \right] \right\}
      \\
  & \times & \braopket{\alpha_4 \alpha_2}{ {\cal T} \exp \left\{
      i \frac{g \mubohr}{2 \hbar} \int_t^{t'}d\tau \left[
        \vec{B}(\vec{R}(\tau))\cdot\bsigma_1
      - \vec{\tilde{B}}(\vec{R}^{\pm}(\tau))\cdot \bsigma_2\right]
      \right\}}{ \alpha_3 \alpha_1}
       \nonumber
 ,
\end{eqnarray}
where $\vec{R}^{-}(\tau) = \vec{R}(t' + t - \tau)$ is 
the time-reversed path of $\vec{R}^{+}\equiv\vec{R}$.

For explicit
evaluation it is convenient to transform 
 this path-integral representation into a differential equation. 
In the case of the diffuson we first have to eliminate the time-reversed
 paths.
As a result of reverting the time integration,
an additional sign appears in the second term of the electromagnetic 
vector potential.
For the Zeeman interaction we can use the relation
\begin{eqnarray}
& & \braopket{\alpha_2}{ {\cal T} \exp \left\{
    -i \frac{g \mubohr}{2 \hbar} \int_t^{t'}d\tau 
   \, \vec{\tilde{B}}(\vec{R}^{-}(\tau)) \bsigma \right\} }
   {\alpha_1}
 =  
\braopket{\alpha_1}{ {\cal T} \exp \left\{
   i \frac{g \mubohr}{2 \hbar} \int_t^{t'}d\tau 
   \, \vec{\tilde{B}}(\vec{R}(\tau)) \bsigma \right\} }
   {\alpha_2}^{*}
   \nonumber \\
& & \qquad \qquad =  
\braopket{\alpha_1}{ {\cal T} \exp \left\{
   -i \frac{g \mubohr}{2 \hbar} \int_t^{t'}d\tau 
   \vec{\tilde{B}}(\vec{R}(\tau)) \bsigma^{*} \right\} }
   {\alpha_2} 
 .
\label{revZeemanPath}
\end{eqnarray}

The latter equation can be proven by writing the time-ordered product 
as a Dyson series and by inserting a resolution of unity in spin space 
 between all products
 $(\vec{B}(x_j)\bsigma)(\vec{B}(x_{j+1})\bsigma)$,
thereby arriving at an expression with terms of the form
$\braopket{\alpha}{B_i(x_j)\sigma_i}{\beta}^{*}$.
Such terms are the complex conjugate of Pauli matrix elements multiplied
by the real number $B_i(x_j)$.
So we can rewrite them as $\braopket{\alpha}{B_i(x_j)\sigma_i^{*}}{\beta}$,
remove the previously inserted unities, and go back to the time-ordered 
product.

Now we can give the differential equations for the propagators
\begin{eqnarray}
\label{diffEqnT}
& & \Bigglb( \frac{\partial}{\partial t'}
     + D\left[-i\frac{\partial}{\partial\vec{x'}}
      -\frac{e}{\hbar} \left[\vec{A}^{\rm em}(\vec{x'}) \pm
       \vec{\tilde{A}}^{\rm em}(\vec{x'}) \right] \right]^2
    \nonumber\\
& & \qquad \qquad -i \frac{g \mubohr}{2\hbar}\left[
     \vec{B}(\vec{x'})\cdot \bsigma_1 
     -\vec{\tilde{B}}(\vec{x'}) \cdot \bsigma_2^{(*)}\right] \Biggrb)
    \hat{\chi}^{C/D}(\vec{x'},\vec{x};t',t)
     = \delta(\vec{x'}-\vec{x})\delta(t'-t)\hat{1\!\!1},
\end{eqnarray}
where $\hat{\chi}^{C/D}(\vec{x'},\vec{x};t',t)$ is a matrix in four-dimensional
spin space. 
The upper sign is for the cooperon~\cite{LSG}, the lower sign and the complex
conjugate of $\bsigma_2$ for the diffuson.
Passing to Fourier space and operator notation, the above equation becomes 

\begin{equation}
\label{diffEqnOmega}
\left( i \omega 
     - D\frac{(2\pi)^2}{L^2} h^{C/D} 
    \right)
    \hat{\chi}^{C/D}_\omega
     = \hat{1\!\!1},
\end{equation}
where the effective Hamiltonian $h^{C/D}$ is defined 
 in Eq.~(\ref{hamiltonian}).

Finally we express the conductance correlation in terms of
the operators $\chi^{C/D}_\omega$.
We note that with 
$\chi^C_{\alpha_1\alpha_2,\alpha_3\alpha_4}(\vec{x},\vec{x'},\omega)^{*} =
\smbraopket{\vec{x'},\alpha_4\alpha_2}{\hat{\chi}^C_\omega}
    {\vec{x},\alpha_3\alpha_1}^{*} = 
\smbraopket{\vec{x},\alpha_3\alpha_1}{\hat{\chi}^{C\dagger}_\omega}
    {\vec{x'},\alpha_4\alpha_2}
$
, and
$\chi^D_{\alpha_1\alpha_2,\alpha_3\alpha_4}(\vec{x},\vec{x'},\omega)^{*} =
\smbraopket{\vec{x'},\alpha_4\alpha_1}{\hat{\chi}^D_\omega}
    {\vec{x},\alpha_3\alpha_2}^{*} = 
\smbraopket{\vec{x},\alpha_3\alpha_2}{\hat{\chi}^{D\dagger}_\omega}
    {\vec{x'},\alpha_4\alpha_1}
$
we can simplify the terms in Eq.~(\ref{ucfChi}):

\begin{equation}
  \int d\vec{x} \, d\vec{x'} \sum_{\alpha_1,\ldots,\alpha_4} 
   \left|\chi^{C/D}_{\alpha_1\alpha_2,\alpha_3\alpha_4}(\vec{x},\vec{x'},\omega)\right| ^2 
 = {\rm Tr}\, \hat{\chi}^{C/D}_\omega \hat{\chi}^{C/D\dagger}_\omega
,
\end{equation}
and

\begin{equation}
  \int d\vec{x} \, d\vec{x'} \sum_{\alpha_1,\ldots,\alpha_4} 
  \chi^{C/D}_{\alpha_1\alpha_2,\alpha_3\alpha_4}(\vec{x},\vec{x'},\omega) 
  \chi^{C/D}_{\alpha_2\alpha_1,\alpha_4\alpha_3}(\vec{x'},\vec{x},\omega) 
 = {\rm Tr}\, \hat{\chi}^{C/D}_\omega \hat{\chi}^{C/D}_\omega
.
\end{equation}

\section{Finite Temperature Integrals}

\label{appMatsubara}

We shall explain here the integrations performed to obtain
 Eq.~(\ref{eqnUCFTfinite}).
We are interested in 
\begin{equation}
\label{eqnMatsubaraIall}
I = \int d \epsilon' n'(\epsilon') J
  = \int d \epsilon' n'(\epsilon') \int d\epsilon \, n'(\epsilon) 
    \left( \frac{1}{d}\frac{1}{(\epsilon-\epsilon' + \miAl)^2 + \miDel^2}
    +2 {\rm Re}\frac{1}{(i\epsilon-i\epsilon' + i \miAl - \miDel )^2} \right)
\end{equation}
 with $a,c$ real and $c>0$.
We consider a rectangular integration contour $\Gamma$ with one side lying on
 the real axis, extending $M = 2\pi l/\beta$ towards the positive imaginary
 axis and the same amount on each side of the real axis.
For any positive integer $l$, the absolute value of the
 Fermi function is bounded above on such a contour:
 $ |n(z)| \Big|_{\Gamma}<2$.
The integrands considered further below are a product of the Fermi function
 and a rational function decaying with at least $|z|^{-2}$.
The integral of
 these products over the section of $\Gamma$ in the upper half plane,
 will thus vanish for $M\to\infty$, as we have $|z|\geq M$ on this contour.
We further note, that the complex expansion of the Fermi function $n(z)$
 has its poles at $z=i\omega_n$, where $\omega_n = \pi n/\beta$ are the
 Matsubara frequencies and $n$ is an odd integer.

We expand the first rational function in Eq.~(\ref{eqnMatsubaraIall}) 
 into partial fractions and then integrate $J$ by parts:
\begin{eqnarray}
\nonumber
J &=& \int d \epsilon n(\epsilon) \left\{
   \frac{1}{d} \frac{1}{2 i \miDel}
   \left(   \frac{1}{(\epsilon - \epsilon' + \miAl - i \miDel)^2}
   -  \frac{1}{(\epsilon - \epsilon' + \miAl + i \miDel)^2}  \right)
  + 2 {\rm Re} \frac{-2}{(\epsilon - \epsilon' + \miAl + i \miDel)^3} \right\}
\\
\label{eqnIpartInt}
 &=& {\rm Re} \int d \epsilon n(\epsilon) \left\{
   \frac{1}{d} \frac{i}{\miDel (\epsilon - \epsilon' + \miAl + i \miDel)^2}
   -\frac{4}{(\epsilon - \epsilon' + \miAl + i \miDel)^3}
   \right\}
 .
\end{eqnarray}
We now evaluate the integral along the contour described above.
As the poles of the rational functions in Eq.~(\ref{eqnIpartInt}) are
 in the lower half plane at $\epsilon' - \miAl -i \miDel$, they are
 not within the integration contour.
Applying Cauchy's residue theorem and accounting for 
 the residues of the Fermi function
 ${\rm res}\, n(z)|_{z=i\omega_n}=(-1/\beta)$ yields
\begin{equation}
 J = \frac{2\pi}{\beta} {\rm Re} \sum_{n \, \rm{odd}>0} \left\{
   \frac{1}{d} \frac{1}{\miDel (i\omega_n - \epsilon' + \miAl + i \miDel)^2}
   +\frac{4i}{(i\omega_n - \epsilon' + \miAl + i \miDel)^3}
   \right\}
  .
\end{equation}

For the second integration in Eq.~(\ref{eqnMatsubaraIall}), 
 we replace the expression in braces in the above equation by
 its complex conjugate.
As before, we first integrate by parts over $\epsilon'$ and apply
 the residue theorem afterwards. This results in

\begin{eqnarray}
I &=& \frac{4\pi}{\beta} {\rm Re} \sum_{n \, \rm{odd}>0}
   \int d \epsilon' n(\epsilon') \left\{
    \frac{1}{d} \frac{1}{\miDel (i\omega_n + \epsilon' - \miAl +i \miDel)^3}
   +\frac{6i}{(i\omega_n + \epsilon' - \miAl + i \miDel)^4}
 \right\}
\nonumber \\
\label{eqnMatsubaraIallSol}
 & = & 
  \frac{8\pi^2}{\beta^2} {\rm Re} \sum_{n,m \, \rm{odd}>0}  \left\{
    \frac{1}{d} \frac{1}{\miDel (\omega_n + \omega_m + i\miAl + \miDel)^3}
   +\frac{6}{(\omega_n + \omega_m + i\miAl + \miDel)^4}
 \right\}.
\end{eqnarray}

\ifpreprintsty\else\bmcsep\begin{multicols}{2}\fi

\section{UCFs $\dgUCF_\homtag$ for homogeneous fields}
\label{secUcfLimits}

For homogeneous fields we have $\eta=\tilde{\eta}=0$ and $f=0$,
 thus the hamiltonians $h^{C/D}$ [Eq.~(\ref{eqnHMatrixElement})]
 become diagonal with the matrix elements
 $j^2 + i \alpha \kappa - i \tilde{\alpha} \tilde{\kappa}$.
Now we evaluate the propagators $\hat{\chi}^{C/D}$ [Eq.~(\ref{eqnChi})]
 and by evaluating 
 the integrals over the Fermi functions in Eq.~(\ref{eqnUcfTrChi})
 explicitly by using standard Matsubara techniques, as explained
 in App.~\ref{appMatsubara}.
We obtain $\dgUCF_\homtag = \dgUCF_{\homtag,\,C} + \dgUCF_{\homtag,\,D}$, 
 where
\begin{eqnarray}
\label{eqnUCFFull}
\dgUCF_{\homtag,\atop C/D} && = 
 \left(\frac{e^2}{h}\right)^2 \! \frac{1}{8\pi^6} 
  \left(\frac{L^2}{L_T^2}\right)^2 {\rm Re} \, 
   \sum_{\alpha, \tilde{\alpha} = \pm 1} 
   \sum_{j=-\infty}^\infty 
 \left. \sum_{n,m} \right.'
\nonumber \\ && 
  \left\{
   \frac{1}
    {d \, (\gamma^{C/D}\!+j^2) 
  \left[\bnm + \gamma^{C/D}\!+j^2 
         \! +\! i(\alpha\kappa - \tilde{\alpha}\tilde{\kappa})  \right]^3 }
\right. \nonumber \\ & & \quad \left.  %noPP
 + \frac{6}
  {\left[\bnm + \gamma^{C/D}\!+j^2 \! +\! i(\alpha\kappa - \tilde{\alpha}\tilde{\kappa})  \right]^4}
  \right\}
 ,
%\nonumber \\
\end{eqnarray}
and we have introduced $\bnm = (n+m) (L/L_T)^2/4\pi$.
Here $n$ and $m$ are positive, odd integers.
The Aharonov-Bohm flux is implemented by replacing
 $j \to j - (\phi/\phi_0 \pm \tilde{\phi}/\phi_0)$.
For further evaluation we now set $\kappa = \tilde{\kappa}$.
We describe the summation of cooperon and diffuson terms with
 a prefactor $\beta$,
which is $1$ if both terms contribute and $2$ if time-reversal
symmetry is broken, so the cooperon contribution vanishes.
Thus we have
 $\dgUCF_\homtag
  \approx (2/\beta)\dgUCF_{\homtag,\,D}$
and from now on we only consider the dephasing parameter
 $\gamma = \gamma^D = L^2/(2\pi L_\varphi)^2$, according to Eq.~(\ref{eqnLCD}).

If the spin-channel mixing is suppressed (i.e.\ $\kappa \gg \gamma$)
 in Eq.~(\ref{eqnUCFFull}), we
can replace the sum over the spins $\sum_{\alpha\tilde{\alpha}}$
by the number of spin states $g_s$.
For weaker magnetic fields ($\kappa \ll \gamma$) we have
full spin degeneracy and obtain the factor $g_s^2$.
Accounting for valley degeneracy yields a factor $g_v^2$.

Since we will check our results against the ones given in 
 Ref.~\onlinecite{beenakker}, where one-dimensional systems 
 were considered, we take $d=1$.
Since we will evaluate some limiting cases below,
 where $L \gg 2\pi L_\varphi$,
 we have $\gamma \gg 1$ and can therefore replace the $j$-sum 
 in Eq.~(\ref{eqnUCFFull}) by an integral.
The Aharonov-Bohm phase can then be removed 
 by shifting the integration variable $j$
 and we obtain
\begin{eqnarray}
\label{eqnUCFadApprox}
\dgUCF_\homtag & = & 
  \left(\frac{e^2}{h}\right)^2 \! \frac{1}{8\pi^6} \frac{2 g_s^2 g_v^2}{\beta}
  \left( \frac{L^2}{L_T^2} \right)^2
   \left. \sum_{n,m} \right.' 
   \int_{-\infty}^{\infty} dj
\\ \nonumber && 
  \left\{  \frac{1}
    {d \, (\gamma + j^2)
\left[\bnm +  \gamma + j^2 \right]^3 } \right.
\left.  + \frac{6}{
\left[\bnm +  \gamma + j^2 \right]^4 }
  \right\}.
\end{eqnarray}
In the limit $(2\pi)^2 L_\varphi^2 \ll L^2, 2 \pi L_T^2$, we have

\begin{equation}
\frac{2\pi L_T^2}{L^2} \, (\gamma+j^2)
 \geq 
 \frac{L_T^2}{2\pi L_\varphi^2} \gg 1.
\end{equation}
Thus, we can use Poisson's summation formula to replace the summation over
$n$ and $m$ in Eq.~(\ref{eqnUCFadApprox}) by integration to arrive at

\begin{eqnarray}
\label{eqnUCFHomLimit1}
\dgUCF_\homtag  &=& 
   \frac{3}{4\pi^4} 
   \frac{g_s^2 g_v^2}{\beta} \left(\frac{e^2}{h}\right)^2
   \int_{-\infty}^{\infty}
   \frac{dj}{\left(\gamma + j^2 \right)^2}
\nonumber \\ & &  %noPP
=  3 \, \frac{g_s^2 g_v^2}{\beta}   \left(\frac{e^2}{h}\right)^2
   \left(\frac{L_\varphi}{L}\right)^3
 .
\end{eqnarray}

We now consider another limit, $2\pi L_T^2 \ll L^2, (2\pi)^2 L_\varphi^2$.
Again, we start from Eq.~(\ref{eqnUCFadApprox}), 
 but now we first calculate the integral over $j$,
 which has the dominant contribution $\pi \gamma^{-1/2} b^{-3}$
 since $1 \ll \gamma \ll \bnm$.
Thus we obtain
\begin{eqnarray}
\label{eqnUCFHomLimit2}
\dgUCF_\homtag & = & 
  \frac{4}{\pi}\frac{g_s^2 g_v^2}{\beta}
  \left(\frac{e^2}{h}\right)^2
  \frac{L_T^2}{L^2}  \frac{L_\varphi}{L} 
   \left.\sum_{n,m}\right.' \frac{1}{\left(\frac{1}{2}(n+m)\right)^3}
\nonumber \\ & %noPP
 =
& %noPP 
\frac{2\pi}{3}\frac{g_s^2 g_v^2}{\beta}
  \left(\frac{e^2}{h}\right)^2
  \frac{L_T^2}{L^2}  \frac{L_\varphi}{L}.
\end{eqnarray}

Indeed, our results $\dgUCF_\homtag$ given in 
 Eqs.~(\ref{eqnUCFHomLimit1}) and~(\ref{eqnUCFHomLimit2})
 agrees with these of Ref.~\onlinecite{beenakker}. 
Thus, on the one hand, we have confirmed that
 the result from Ref.~\onlinecite{LSG} [used in Eq.~(\ref{eqnUcfTrChi})]
 is consistent with earlier calculations.\cite{lee,altshuler,beenakker}
On the other hand,
 in Eq.~(\ref{eqnUCFFull}) we have given an explicit
 formula (not known before as far as we are aware of)
 describing how the spin-channel mixing becomes suppressed for
 increasing magnetic fields, 
 such that $\dgUCF_\homtag$ contains a
 prefactor $g_s^2$ for low and $g_s$ for high magnetic fields.

\ifpreprintsty\else\clearpage\end{multicols}\fi


\begin{references}
\bibitem{footnoteGeneralize}
Implementing magnetic textures with different winding numbers 
$f$ and $\tilde{f}$ can be easily done by replacing the coefficient
of  $\sigma_{2z}$ in Eq.~(\ref{eqnJCD}) with  $\pm\tilde{f}/2$.
This will result in different diagonal elements in the matrices given below.
As we have verified,
 within this approach
 the only further generalization of the magnetic texture is an
 arbitrary rotation axis of the field.\cite{engelDipl}
Otherwise 
 it is not possible to find a
 linear combination of $\partial_x$ and Pauli matrices
 which commutes with $h^{C/D}$.
\bibitem{fnPeriodicity}
As every quantity $\sum_{j=-\infty}^{\infty} f(j+\Phi)$ is periodic
 in $\Phi$ with period one,
we see that $\dgUCF$ has a periodic dependence on
 the Aharonov-Bohm fluxes $\phi$, $\tilde{\phi}$.
\bibitem{gtte}
The full line in the power spectrum is provided as a guide to the eye.
 It was obtained by zero-padding the data before calculating the DFT.\cite{nr}
\end{references}
\end{document}